\documentclass[10pt,journal]{IEEEtran}

\usepackage[T1]{fontenc}
\usepackage{lmodern}
\usepackage[nopatch=eqnum]{microtype}
\usepackage{graphicx}
\usepackage{booktabs}
\usepackage{tabularx}
\usepackage{array}
\usepackage{makecell}
\usepackage{multirow}
\usepackage{enumitem}
\usepackage{cite}
\usepackage{url}
\usepackage[hidelinks]{hyperref}
\usepackage[nameinlink,noabbrev]{cleveref}
\usepackage{tikz}
\usetikzlibrary{arrows.meta,positioning,fit,calc}
\usepackage[dvipsnames]{xcolor}
\usepackage{threeparttable}
\usepackage{longtable}
\usepackage{ragged2e}

\newcommand{\institution}{Purdue University}
\newcommand{\department}{the Department of Computer Science}
\newcommand{\prerequisitecourse}{CS~25100, Data Structures and Algorithms}
\newcommand{\CAPcourse}{Purdue University's ENE 50600, Content, Assessment and Pedagogy}
\newcommand{\CAPauthor}{Ruth A. Streveler}

\newcolumntype{Y}{>{\RaggedRight\arraybackslash}X}
\newcolumntype{P}[1]{>{\RaggedRight\arraybackslash}p{#1}}
\setlist[itemize]{leftmargin=*,itemsep=1pt,topsep=2pt}
\setlist[enumerate]{leftmargin=*,itemsep=1pt,topsep=2pt}

\title{Removing the Competition from\\Introductory Competitive Programming:\\
A Backward-Designed, Mastery-Oriented\\Curriculum Sequence}

\author{Ethan Dickey%
\thanks{Ethan Dickey is with the Department of Computer Science, Purdue University, West Lafayette, IN, USA (dickeye@purdue.edu).}}
\hypersetup{
  pdftitle={Removing the Competition from Introductory Competitive Programming: A Backward-Designed, Mastery-Oriented Curriculum Sequence},
  pdfauthor={Ethan Dickey},
  pdfkeywords={competitive programming, curriculum design, backward design, constructive alignment, mastery learning, active learning, oral assessment, contest preparation}
}

\begin{document}
\maketitle

\begin{abstract}
Competitive programming courses often inherit the visible features of programming contests -- timed rounds, rankings, penalty systems, and topic lists -- before articulating what novices should learn. This curriculum design case argues that competitive programming tasks and competitive performance are related but distinct educational objects. Introductory courses should first develop algorithmic problem-solving competence: recognizing problem structure, selecting and combining paradigms, decomposing tasks, analyzing complexity, implementing and debugging solutions, and explaining the reasoning that produced them. Contest-performance competence -- rapid triage, speed under uncertainty, strategic risk management, and team communication -- should be introduced only after that foundation is sufficiently broad.

The paper reconstructs a multi-year redesign of a three-course competitive programming sequence using an engineering-education framework that aligns content, assessment, and pedagogy through backward design. The resulting CP1 and CP2 courses use mastery-oriented online-judge problem sets, a breadth requirement, sampled oral code interviews, collaborative in-class reasoning, paired problem solving, fading scaffolds, and metacognitive reflection. A subsequent CP3 course restores timed individual and team contests as the authentic capstone environment. Competition thus enters the sequence through an explicit pedagogical dosage and sequencing decision.

This article is a transparent design case rather than an efficacy trial. It reports the rationale, artifacts, implementation evolution, design tradeoffs, and a tiered evaluation agenda without claiming causal learning gains. It contributes a two-layer competence model, an aligned curriculum architecture, implementation-ready assessment protocols, and an adapted course-design template for competitive programming educators.
\end{abstract}


\section*{Funding}
No external funding was received for this work.

\section*{Competing Interests}
The author declares no competing interests.

\section{Introduction}
\label{sec:intro}

Competitive programming (CP) is simultaneously a class of algorithmic tasks and a social-performance format. The tasks present precisely specified computational problems, enforce time and memory bounds, and provide objective evaluation through hidden tests. The format adds scarcity and comparison: a fixed contest window, penalties, rank ordering, strategic problem selection, and, in team events, rapid communication. These two meanings are usually bundled together. A course called ``competitive programming'' is therefore often assumed to require competition from the first week.

That assumption is pedagogically consequential. CP courses are frequently assembled from canonical topics and problems drawn from textbooks, online judges, or the prior contest experience of an instructor or student coach \cite{skiena1999training,halim2020cp4,laaksonen2020guide,skiena2003challenges}. Many such courses are valuable, and a growing literature documents contest ecosystems, curricular practices, platforms, and educational benefits \cite{wasik2018taxonomy,yuen2023competitive,bandeira2019teaching,guha2025everybody,lim2026paradigm}. Yet the field has devoted less attention to a prior design question: \emph{Which parts of competitive programming should be taught together, and in what developmental order?}

The redesign reported here began from a practical problem at \institution. A popular elective sequence had evolved through instructor and teaching-assistant judgment, but its topic order, assessments, and classroom activities had not been derived from explicit enduring outcomes. A 2021 content review identified duplicated material, abrupt difficulty changes, and dependencies taught in reverse order; for example, answer-space bisection appeared before the greedy feasibility reasoning on which many bisection problems depend \cite{dickey2021content}. The course also faced a familiar tension: it was marketed partly as interview and contest preparation, but students needed a much broader repertoire of classification, decomposition, analysis, implementation, and debugging skills before timed performance could be a valid representation of learning.

The redesign used a graduate engineering-education course framework that treats curriculum design as an integrated engineering problem: specify educational requirements, justify them with evidence, design aligned assessments and pedagogy, prototype the course, and iterate \cite{biggs1996alignment,wiggins2005ubd,fink2013significant,hansen2011idea,strevelersmith2020cap}. The resulting sequence makes a deliberately provocative move: it removes most consequential competition from the introductory courses while retaining the defining task ecology of CP. Students still solve authentic CP problems against hidden tests and resource constraints. What is deferred is the demand to perform that work rapidly, rank against peers, and coordinate under contest pressure. Those demands return in an advanced course designed explicitly for contest preparation.

We do not claim that competition is educationally harmful or inauthentic. Structured competition can motivate, focus practice, and support excellence, but its effects depend on goals, learners, climate, and implementation \cite{tauer2004competition,murayama2012achievement,raman2018motivation,lim2026paradigm}. Instead, we claim that \emph{competition is a curriculum variable, not a default}. For mixed-experience introductory cohorts, timed rank competition can combine too many partially developed capabilities into one noisy performance. A mastery-oriented sequence can establish the schemas and self-regulatory routines that make later contest performance more learnable and more interpretable.

This paper makes four contributions:
\begin{enumerate}
    \item It distinguishes \emph{algorithmic problem-solving competence} from \emph{contest-performance competence} and uses that distinction to justify deferred competition.
    \item It presents a backward-designed CP1--CP2--CP3 sequence that aligns enduring outcomes, learning objectives, assessments, and active-learning pedagogy.
    \item It documents implementation mechanisms -- including mastery-oriented online-judge assignments, breadth gates, sampled oral interviews, collaborative problem solving, and a bounded generative AI intervention -- together with their boundary conditions and design tradeoffs.
    \item It provides an evaluation agenda and appendices that other educators can adapt without mistaking a design rationale for causal evidence.
\end{enumerate}

\section{Prior Work and Theoretical Commitments}
\label{sec:prior}

\subsection{Competitive Programming as an Educational Environment}

Online judges have made CP tasks widely available and have standardized objective evaluation, immediate verdicts, and performance constraints \cite{wasik2018taxonomy,mirzayanov2020codeforces,diluigi2016oii}. CP has been used in dedicated courses, introductory programming, enrichment programs, and contest-team preparation \cite{ribeiro2008early,combefis2012novel,bandeira2019teaching,guha2025everybody}. Reported benefits include sustained practice, engagement, algorithmic reasoning, computational thinking, and independent problem solving \cite{yuen2023competitive,raman2018motivation}. Systems for personalized problem recommendation and gamification further demonstrate that the learning environment can be engineered rather than treated as a static list of problems \cite{dimascio2018personalized,diluigi2016oii}.

The literature also differentiates course purposes. Some work foregrounds contest strategy, team preparation, and performance under time pressure \cite{trotman2008strategy,bloomfield2016guide,luo2025contest}. Other work embeds CP-style tasks within programming education or emphasizes practice and problem-solving development \cite{ribeiro2008early,bandeira2019teaching,yuen2023competitive}. Cooperative and competitive structures can coexist; collaborative behaviors can be designed even inside contest-oriented assignments \cite{gonzalez2019collaborative,tauer2004competition}. This diversity is a strength, but it also means that the label \emph{competitive programming course} is under-specified. A contest-team practicum and a first encounter with applied algorithmic problem solving should not be expected to share the same outcomes or assessment conditions.

Recent work makes the design contrast especially visible. Luo proposes a contest-based curriculum in which realistic time pressure is an explicit formative component \cite{luo2025contest}. That design addresses an authentic need: learners who already possess broad algorithmic knowledge must practice performing under contest constraints. The present design is complementary rather than contradictory. It specifies the developmental stage that should precede that approach and locates contest-based assessment primarily in CP3. Similarly, analyses of performance consistency across contests show that contest structures sample different dimensions of skill and are not interchangeable measures \cite{luodickey2026consistency}. This supports treating the contest format itself as an object of alignment rather than a neutral container.

\subsection{Backward Design and Constructive Alignment}

Constructive alignment requires that intended learning outcomes, learning activities, and assessments all target the same forms of knowledge and performance \cite{biggs1996alignment}. Backward design begins with desired understanding and transfer, identifies acceptable evidence, and only then selects learning experiences \cite{wiggins2005ubd}. Fink and Hansen similarly emphasize integrated course design, significant learning, big ideas, and authentic assessment \cite{fink2013significant,hansen2011idea}. The CAP framework used in this redesign operationalizes these principles through explicit attention to content, assessment, and pedagogy, together with learner context, difficult concepts, evidence, and alignment \cite{strevelersmith2020cap}, \cite{streveler2021capsyllabus}.

This approach matters in CP because a topic list is not a curriculum. ``Dynamic programming, graphs, strings, geometry'' specifies content coverage but not what students should notice, explain, transfer, or produce. Nor does a count of accepted submissions establish whether students can classify an unfamiliar problem, justify an algorithm, or reproduce their own work. The redesign therefore treats canonical algorithms as resources for developing higher-level capabilities rather than as the terminal outcomes themselves.

\subsection{Learning Commitments}

Six commitments connect the design to learning research.

First, \textbf{complex problem solving should be scaffolded and sequenced}. Novices and intermediates possess fewer organized schemas and can expend substantial working-memory resources on surface features, syntax, and local implementation details \cite{sweller1988load,vanmerrienboer2005load,robins2003learning}. Adding time scarcity, ranking, and strategic triage before core schemas are stable may increase task demands without increasing the learning value of the target practice. Worked examples, coaching, and fading support are therefore used early, with autonomy increasing over time \cite{wood1976scaffolding,collins1989apprenticeship}.

Second, \textbf{competence develops through repeated, distributed, goal-directed practice with feedback} \cite{ericsson1993practice,cepeda2006distributed}. Mastery-learning traditions support allowing additional attempts when the purpose is eventual competence rather than ranking students by the speed of initial acquisition \cite{bloom1968mastery,kulik1990mastery}. The weekly problem sets retain strict correctness and efficiency standards while allowing repeated submissions before the deadline.

Third, \textbf{automated feedback is useful but partial}. Online judges provide rapid behavioral evidence about output correctness and resource use, but their verdicts reveal little about conceptual understanding, reasoning quality, authorship, or code comprehensibility \cite{alamutka2005automatic,keuning2019feedback}. Effective feedback should help learners answer where they are going, how they are progressing, and what to do next; a bare ``wrong answer'' or ``time limit exceeded'' supports only part of that process \cite{hattie2007feedback,shute2008feedback,butlerwinne1995feedback}. Judge feedback is thus accompanied by human dialogue and reflection.

Fourth, \textbf{learning activities should move beyond passive exposure}. The ICAP framework predicts stronger learning when students construct and interact with ideas rather than merely receive them \cite{chi2009active,chiwylie2014icap}. Active and cooperative learning have broad support in STEM and engineering education \cite{prince2004active,freeman2014active,springer1999smallgroup,johnson1998cooperative}. Based on this, we foreground think-pair-share, public comparison of solution ideas, and paired implementation, rather than treating them as supplementary.

Fifth, \textbf{metacognition and transfer must be made explicit}. Problem-solving expertise includes monitoring, strategy selection, and reflection on why a solution path worked \cite{flavell1979metacognition,schoenfeld1985problem,polya1945solve,zimmerman2002srl}. Transfer is more likely when learners abstract structure across contrasting cases and are prepared to learn from new situations \cite{bransford1999transfer}. Students are repeatedly asked to name the paradigm, identify decisive constraints, explain the brute-force baseline, and articulate the key insight that made an efficient solution possible.

Sixth, \textbf{motivation and participation structures should not assume that all students experience competition identically}. Competition can increase performance through approach motivation while also increasing avoidance, anxiety, or disengagement for some learners \cite{murayama2012achievement}. Autonomy support, attainable competence, meaningful choice, and relatedness are therefore important design conditions \cite{ryandeci2020sdt,reeve2009autonomy}. These conditions are especially relevant in computing environments where prior experience and belonging are unevenly distributed \cite{margolis2002clubhouse}.

\section{Design Case Method and Context}
\label{sec:method}

\subsection{Article Type and Design Questions}

This article is a retrospective curriculum design case. Design cases communicate situated design knowledge by making goals, constraints, decisions, artifacts, tradeoffs, and revisions visible \cite{boling2010designcases}. They are related to, but distinct from, design-based research: this paper does not claim a controlled intervention, causal effect, or fully instrumented iterative study \cite{dbr2003design,mckenney2018edr}. Its evidentiary contribution is a transparent and theoretically grounded reconstruction of a curriculum design that has been implemented and revised.

Three design questions organize the case:
\begin{enumerate}[label=\textbf{DQ\arabic*:},leftmargin=*]
    \item What enduring capabilities should define introductory competitive programming when contest performance is not assumed to be the immediate objective?
    \item How can content, assessment, and pedagogy be aligned so that students repeatedly practice and demonstrate those capabilities?
    \item Where, and under what assessment conditions, should timed competition enter a multi-course sequence?
\end{enumerate}

\subsection{Institutional Setting and Constraints}

The sequence was developed in \department{} at \institution. CP1 currently enrolls approximately 90--100 students per offering. CP2 enrolls approximately 20--30 students, and CP3 enrolls approximately 20--30 students when it is offered every other semester. The formal prerequisite for CP1 is \prerequisitecourse{}, with instructor approval available for students who can demonstrate equivalent preparation. The sequence therefore begins after students have encountered fundamental data structures, graph traversal, sorting, and asymptotic analysis, but before most have developed a broad competitive programming algorithmic repertoire.

CP1 and CP2 were initially approved as two-credit, twelve-week courses with approximately 110 minutes of class time per week; the implementation described here later used two 75-minute meetings each week. CP3 follows the same compressed-course model when offered. These variations are important design constraints: the architecture must support large enrollment and substantial weekly practice without relying on conventional exams or an extensive laboratory component \cite{dickey2022cp1proposal},\cite{dickey2022cp2proposal},\cite{dickey2023cp3proposal},\cite{dickey2026syllabus}.

The courses serve overlapping purposes in interview preparation, applied algorithmic problem solving, and ICPC participation. They use online-judge infrastructure, permit multiple implementation languages, and draw selectively from a canonical competitive programming text. Students enter with sharply different prior experience: some have competed extensively, while others know only the material expected from the prerequisite course. The design must therefore remain rigorous for experienced students, legible to newcomers, and scalable enough to support frequent feedback in a cohort near 100.

\subsection{Artifact Corpus and Analytic Procedure}

The design record comprises the artifacts summarized in \Cref{tab:artifacts}. They serve as an audit trail for the sequence: they establish when a decision entered the design, how it was implemented, and how it changed across offerings. Most were written by the same practitioner who designed and taught the courses. They are therefore cited for provenance and implementation detail rather than treated as independent authority for the recommendations developed in this article. The warrant for those recommendations comes from the alignment argument, the literature in \Cref{sec:prior}, and the explicitly bounded practitioner knowledge reported here.

The analysis proceeded in three passes. First, the artifacts were ordered chronologically to reconstruct the path from the 2021 content critique to the integrated CAP design, official course proposals, repeated implementation, and the current syllabi. Second, outcomes, assessments, and pedagogies were mapped to identify both alignments and gaps. Third, each consequential design choice was reconsidered against the learning literature and the current operating context. This last pass produced the design tradeoffs in \Cref{sec:evolution}: some practices are proposed as central features, some are retained with proportionate safeguards, and some remain hypotheses for future evaluation.

\begin{table*}[t]
    \centering
    \caption{Curriculum artifact corpus used in the design reconstruction. Dates indicate the design snapshot, not necessarily the date of every later edit.}
    \label{tab:artifacts}
    \footnotesize
    \begin{tabularx}{\textwidth}{P{0.16\textwidth} P{0.17\textwidth} Y Y}
        \toprule
        \textbf{Artifact} & \textbf{Approx. date} & \textbf{Design role} & \textbf{Evidence used in this paper} \\
        \midrule
        Content revision proposal \cite{dickey2021content} & Fall 2021 & Diagnosis of legacy content and proposed CP1/CP2 sequences & Topic dependencies, duplicated material, difficulty discontinuities, candidate problems \\
        CAP course syllabus/template \cite{streveler2021capsyllabus} & Fall 2021 & Design requirements for the ENE curriculum project & Big ideas, outcomes, learner analysis, assessment, pedagogy, alignment, evidence requirements \\
        Complete redesign document \cite{dickey2022design} & April 2022 & First integrated content--assessment--pedagogy specification & Context, enduring outcomes, misconceptions, difficult concepts, objectives, automated assessment rationale \\
        Official CP1 and CP2 proposals \cite{dickey2022cp1proposal}, \cite{dickey2022cp2proposal} & 2022 & Institutional codification of introductory and advanced courses & Catalog scope, prerequisites, learning objectives, problem-set and breadth grading structures \\
        Official CP3 proposal \cite{dickey2023cp3proposal} & 2023 & Contest-preparation capstone design & Timed simulation, teamwork, communication, advanced multi-technique problems \\
        Manuscript notes \cite{dickey2025notes} & May 2025 & Practitioner reflection after multiple offerings & Three-part lesson pattern, interviews, attendance/material policy, participation diagnostic, improvement agenda \\

        Current CP1-CP3 syllabi \cite{dickey2026syllabus}, \cite{dickey2026cp2syllabus}, \cite{ramaswami2026cp3syllabus} & Spring 2026 & Current implementation snapshots & Current outcomes, topic sequences, problem-set and project structures, oral interviews, and individual/team contest conditions \\
        \bottomrule
    \end{tabularx}
\end{table*}

This is a practitioner analysis. It does not include independent coding, student interviews conducted for research, or inferential analysis of course records. The same person designed, taught, and interpreted the curriculum, so the article separates direct descriptions of implementation from broader claims about what the design should accomplish. Its evidentiary contribution is the inspectable chain from domain assumptions to course outcomes, assessments, pedagogy, and subsequent revision; independent outcome evidence remains future work.

\subsection{Ethics and Scope of Claims}

No identifiable student records, grades, submissions, recordings, or survey responses are analyzed in this manuscript. Descriptive statements about the course are drawn from curriculum artifacts, and judgments about implementation are practitioner reflections. Accordingly, the paper does not report human-subjects findings.

The redesigned courses were implemented across nine offerings from Spring 2022 through Spring 2026. That history supports a claim of feasibility and iterative use, not a claim of effectiveness. The Spring 2026 syllabi are treated as design snapshots rather than evidence that every described component produced a particular outcome.

\section{Two Layers of Competitive Programming Competence}
\label{sec:layers}

The phrase \emph{competitive programming skill} compresses at least two competence layers. The first is algorithmic problem-solving competence: understanding a statement, modeling it, identifying structural cues, selecting or composing algorithms, estimating feasibility, implementing a solution, generating tests, debugging, and explaining the result. The second is contest-performance competence: producing that work quickly while allocating scarce time, selecting among problems, managing uncertainty, communicating with teammates, and responding to rank and penalty information.

\Cref{tab:layers} separates the two layers for design purposes. They interact, and expert contestants integrate them seamlessly. Nevertheless, they have different prerequisite structures and are assessed under different validity conditions. A student may understand dynamic programming but be slow at recognizing it; another may recognize a pattern quickly but implement unreliably; a third may solve well alone but communicate poorly on a team. A single contest score conflates these capabilities with problem-set sampling, prior exposure, affective response, and strategic choices.

\begin{table*}[t]
\centering
\caption{A two-layer model of competitive programming competence. The difference is analytical rather than absolute.}
\label{tab:layers}
\footnotesize
\begin{tabularx}{\textwidth}{P{0.18\textwidth} Y Y}
\toprule
 & \textbf{Layer 1: algorithmic problem-solving competence} & \textbf{Layer 2: contest-performance competence} \\
\midrule
Primary question & Can the learner produce, justify, implement, and verify an efficient solution? & Can the learner deploy that competence rapidly and strategically under contest constraints? \\
Knowledge and skills & Problem representation; paradigm recognition; decomposition; data-structure and algorithm selection; complexity analysis; implementation; testing; debugging; explanation; transfer & Triage; time allocation; speed--accuracy calibration; risk management; scoreboard interpretation; penalty strategy; role coordination; concise team communication; emotional regulation \\
Typical evidence & Untimed or generously timed novel problems; oral explanation; trace and test generation; code review; reflection; transfer tasks & Timed individual or team contests; time to first accept; in-contest problem-selection quality; penalties; communication quality; post-contest strategic analysis \\
Major validity threats & Memorized solutions; trial-and-error submissions; opaque authorship; uneven prerequisite knowledge & Problem-set luck; prior contest exposure; anxiety; typing speed; team composition; strategy; rank dependence \\
Best curricular location & CP1 and CP2, with increasing integration and decreasing scaffolding & CP3 and co-curricular team training, after broad Layer 1 competence \\
\bottomrule
\end{tabularx}
\end{table*}

\subsection{Why Defer, Rather Than Eliminate, Competition?}

The case for deferral has four parts.

\textbf{Construct validity.} If CP1 intends to assess classification, decomposition, complexity reasoning, implementation, and debugging, a highly timed contest introduces additional variance. Low scores can reflect missing target competence, but they can also reflect slow reading, unfamiliar syntax, poor triage, anxiety, or a single early bug. Timed performance becomes more defensible after the underlying repertoire is established and the course explicitly teaches contest strategy.

\textbf{Cognitive load and sequencing.} Introductory CP problems are already element-interactive: students must coordinate statement interpretation, mathematical modeling, algorithm choice, data structures, asymptotic constraints, language details, and debugging. Contest mechanics add simultaneous demands. Deferral allows early practice to target the hard conceptual parts before those parts must be executed at speed \cite{sweller1988load,vanmerrienboer2005load,perkins2009whole}.

\textbf{Practice quality.} Deliberate practice requires tasks at the edge of current competence, opportunities to correct errors, and feedback that can be acted upon \cite{ericsson1993practice}. A student who spends a contest stalled on one problem may experience authenticity without sufficient learning density. Weekly, topic-linked, retriable problems make it easier to distribute practice and intentionally sample the desired schemas.

\textbf{Motivational access.} Rank competition can be energizing for students with established competence and a performance-approach orientation, while creating avoidance or belonging threats for students who are still acquiring the local norms \cite{murayama2012achievement,ryandeci2020sdt,margolis2002clubhouse}. A mastery-oriented entry point does not guarantee inclusion, but it reduces the extent to which early public speed becomes the definition of who ``belongs'' in CP.

Deferral also avoids a false dichotomy. CP1 and CP2 remain authentic because students solve real CP tasks with hidden tests and resource limits. What changes is the \emph{performance envelope}. CP3 then makes speed, strategy, and team communication explicit outcomes, rather than treating them as invisible prerequisites. In this sequence, competition is neither removed from the domain nor used as a universal teaching method; it is reserved for the stage at which it becomes an aligned whole-task performance.

\section{Content Design: From Enduring Outcomes to a Three-Course Sequence}
\label{sec:content}

\subsection{Big Ideas and Enduring Outcomes}

The design begins with the competitive programming domain rather than with a weekly topic list. At the domain level, we posit three organizing ideas. First, computational problems usually admit multiple solution paths whose feasibility depends on structure and constraints. Second, CP links algorithmic theory to executable application through compact, testable whole problems. Third, decomposition and solution generation can be practiced systematically instead of being left entirely to intuition. These ideas describe the intellectual work of the domain, not yet the scope of any single course \cite{dickey2022design}.

The next step is to narrow from domain to course. A three-course undergraduate sequence cannot cover every algorithm family, so it should preserve the capabilities that remain useful when individual algorithms are forgotten. We therefore select four enduring capabilities: recognizing problem structure, decomposing problems, designing efficient solutions, and analyzing time and space. Canonical algorithms, data structures, and implementation patterns are important-to-know resources that enable those capabilities, but are deliberately not treated as the terminal outcomes of the sequence.

For assessment and publication, we operationalize this domain-to-course transition through six measurable outcome clusters. They consolidate the eleven objectives used in the current CP1 implementation while making the intended evidence more explicit \cite{dickey2026syllabus}:

\begin{enumerate}[label=\textbf{O\arabic*.}]
    \item \textbf{Classify and justify.} Identify plausible problem families and justify the classification using structural cues, constraints, and counterexamples.
    \item \textbf{Decompose and design.} Transform a problem into solvable subproblems, compare a brute-force baseline with improved alternatives, and assemble a correct algorithm.
    \item \textbf{Analyze feasibility.} Derive time and space complexity, relate the analysis to input bounds, and reject designs that cannot satisfy the resource constraints.
    \item \textbf{Implement and verify.} Translate a design into correct code using appropriate data structures, invariants, and language features.
    \item \textbf{Debug systematically.} Generate discriminating tests from the specification and constraints, localize defects, and revise the model or implementation.
    \item \textbf{Explain, reflect, and transfer.} Reconstruct the reasoning behind a solution, identify the decisive insight, and adapt the method to a structurally related but non-isomorphic problem.
\end{enumerate}

These outcomes combine conceptual, procedural, and metacognitive knowledge in the revised Bloom taxonomy \cite{anderson2001taxonomy}. A student who can reproduce Dijkstra's algorithm but cannot decide when its assumptions hold has procedural knowledge without adequate classification and evaluation. A student who accepts an algorithm because it passed the judge but cannot explain its complexity has behavioral success without sufficient evidence of durable understanding. Conversely, a student who can describe a solution but cannot implement it has not completed the ``whole game'' of CP \cite{perkins2009whole}.

\subsection{Difficult Concepts and a Theory of Difficulty}

We treat three practices as the recurring hard parts of introductory CP: selecting a representation or data structure, generating a solution strategy, and locating errors. Each involves troublesome or hidden knowledge that must be made visible through instruction and assessment rather than assumed to emerge from repeated exposure alone \cite{dickey2022design}.

\textbf{Representation and data-structure choice} are difficult because prior coursework can produce ritualized pairings: queues belong with breadth-first search, heaps with priority queues, arrays with indexed data. CP requires representational flexibility. The same conceptual object may be represented as an array, bitmask, graph, interval set, frequency map, or state-transition system depending on the operations and constraints. Our educational target is thus not a catalogue of structures but conditional knowledge about when each representation makes useful operations cheap.

\textbf{Solution generation} is difficult because experts compress many cues into rapid recognition. Input bounds, monotonicity, optimal-substructure clues, exchange arguments, graph properties, and symmetry become tacit signals. Novices see a story; experts see a search space and invariants. The course makes this hidden game visible through a recurring analysis routine: state the brute-force method, estimate its complexity, identify the bottleneck, inspect constraints and structure, propose a transformation, and justify why the new design preserves correctness.

\textbf{Debugging} is difficult in CP because a failed submission underdetermines the cause. The defect may be in the model, algorithm, proof, complexity estimate, boundary condition, numeric representation, input handling, or code. Debugging literature similarly treats the skill as complex and under-taught \cite{mccauley2008debugging}. The course therefore adds custom-test generation as an explicit outcome rather than assuming that repeated judge submissions will automatically produce systematic debugging competence.

The course also addresses misconceptions about the domain: that CP is only interview preparation; that contest speed is equivalent to computer-science ability; that CP code is a model for production software; that a single language is the exclusive or universally superior choice; or that CP is suitable as a first exposure to programming \cite{dickey2022design}. These misconceptions influence identity and strategy. For example, conflating CP speed with general intelligence can make early struggles feel diagnostic rather than developmental. Explicitly separating Layer 1 and Layer 2 competence provides a curricular response.

\subsection{Sequencing Principles}

The content sequence follows four principles.

\textbf{Dependency before coverage.} Topics are ordered by prerequisite reasoning rather than by tradition. Core paradigms precede advanced applications. Greedy reasoning precedes many answer-space bisection tasks because feasibility checks often depend on a greedy construction. Basic traversal and shortest-path reasoning precede advanced graph structures. The 2021 content critique was especially useful here because it made reverse dependencies and repeated coverage visible \cite{dickey2021content}.

\textbf{Paradigms before special cases.} Complete search, greedy methods, divide and conquer, and dynamic programming form an organizing vocabulary. Students are repeatedly asked which paradigm best describes a solution and why nearby paradigms fail. Algorithms such as binary search, prefix sums, graph traversal, and shortest paths are then connected to these larger modes of reasoning rather than taught as isolated recipes.

\textbf{Spiral toward integration.} Early tasks foreground one intended idea. Later tasks contain distracting surface features or require multiple techniques. CP2 increasingly asks students to compose algorithms across categories, and CP3 makes multi-technique problems central. This progression supports transfer while keeping early practice diagnostically interpretable.

\textbf{Reserve speed for a taught performance layer.} CP1 and CP2 expose students to time and memory constraints as properties of algorithms, but not primarily as scarcity of learner time. CP3 adds timed simulation, strategic problem selection, and team communication because those capabilities appear explicitly in its enduring outcomes \cite{dickey2023cp3proposal}.

\begin{table*}[t]
    \centering
    \caption{Developmental roles of CP1, CP2, and CP3 in the current sequence. Topic examples and learning conditions summarize the Spring 2026 implementations rather than attempting to prescribe a universal sequence.}
    \label{tab:sequence}
    \footnotesize
    \begin{tabularx}{\textwidth}{P{0.075\textwidth} P{0.20\textwidth} P{0.26\textwidth} Y Y}
        \toprule
        \textbf{Course} & \textbf{Primary developmental purpose} & \textbf{Representative content} & \textbf{Dominant learning conditions} & \textbf{Principal evidence} \\
        \midrule
        CP1 & Build a broad, explicit repertoire for recognizing and solving single-technique or lightly integrated problems & Useful data structures; complete search; greedy; divide and conquer; bisection; meet in the middle; foundational dynamic programming; prefix sums and two pointers; graph traversal, components, flood fill, topological sorting, and shortest paths; CP2 introduction through minimum spanning trees and disjoint sets & High scaffolding early; generous time; topic-linked practice; collaborative reasoning; frequent retry and feedback & Correct and efficient solutions; breadth across topics; oral explanation; complexity justification; custom tests \\
        \midrule
        CP2 & Expand CP1 observation skills and build a broader technical repertoire for advanced and increasingly mixed problems & Pruning and perspective; sweep lines and monotonic queues; binary exponentiation, sieves, and extended GCD; combinatorics and inclusion--exclusion; RMQ, Fenwick trees, and introductory segment trees; computational geometry and convex hull; tree DP and DFS order; rolling hashes and tries; bitmask DP and graph problems & Topic-focused lectures and think--pair--share; weekly retriable four-problem sets; reduced scaffolding; random mixed-topic review; collaborative problem authoring & Accepted solutions across at least 11 topic sets; sampled oral explanations; clean-code submissions; a complete competitive-programming problem package; mixed-topic transfer \\
        \midrule
        CP3 & Integrate observation, advanced techniques, implementation, and debugging into individual and team contest performance & Advanced bitmask and SOS DP; linear recurrences; advanced segment trees; monotonicity and optimization; LCA and HLD; game theory with Sprague--Grundy functions; network flow with min-cut and min-cost variants; FFT; KMP and Aho--Corasick; offline methods including CDQ and Mo's algorithm & One topic meeting followed by a 90-minute contest; alternating individual and three-person team contests; ICPC-style resource restrictions and balanced coding roles; individual upsolving and brief reflection & Timed in-class acceptances; individual upsolves; repeated individual and team contest performance; per-problem solve and reflection records \\
        \bottomrule
    \end{tabularx}
\end{table*}

\subsection{CP1: Establishing the Algorithmic Repertoire}

CP1 begins with computational conventions and useful data structures, then moves through complete search, greedy methods, divide and conquer, answer-space bisection, dynamic programming, two-pointer techniques, and foundational graph algorithms \cite{dickey2026syllabus}. This progression differs from a conventional algorithms course in emphasis. Proof and asymptotic analysis remain important, but each idea is repeatedly enacted in a short problem whose statement, constraints, implementation, and hidden tests form a complete design-to-execution cycle.

Our goal here is not encyclopedic coverage. Students learn a compact set of generative questions:
\begin{itemize}
    \item What would the brute-force state space be?
    \item Which constraints rule it out or make it plausible?
    \item Is there exploitable ordering, monotonicity, repeated state, or graph structure?
    \item What information must be retained, and which representation makes the required operations efficient?
    \item What invariant or exchange argument would justify the proposed method?
    \item Which adversarial or boundary cases could falsify the implementation?
\end{itemize}

These questions make the content portable: the exact Kattis problems used can (and do frequently) change without changing the curriculum's conceptual spine.

\subsection{CP2: Integration, Ambiguity, and Transfer}

CP2 shifts from identifying individual paradigms to recognizing advanced categories and designing solutions that span categories \cite{dickey2022cp2proposal}, \cite{dickey2026cp2syllabus}. This transition is central to the sequence. Real CP problems are seldom labeled ``use minimum spanning tree'' or ``use digit dynamic programming.'' Students must decide which parts of a problem are graph structure, which are optimization, which are precomputation, and how the pieces interact.

CP2 therefore increases ambiguity in three ways. First, problem statements include plausible but unproductive cues. Second, assignments mix topics so that recently covered topics cannot determine the intended algorithm. Third, students compare multiple feasible solutions and discuss tradeoffs in implementation risk, constant factors, proof burden, and generality. This is the point at which occasional timed diagnostics can be informative, provided they remain low stakes and are followed by analysis. The objective is to reveal recognition latency and integration bottlenecks, not to convert the course grade into a rank contest.

\subsection{CP3: Reintroducing the Competition}

CP3 explicitly centers contest simulation, teamwork, concise communication, advanced topics, and problems that combine techniques from the full sequence \cite{dickey2023cp3proposal}, \cite{ramaswami2026cp3syllabus}. It is therefore the point at which competition becomes the primary learning environment rather than an incidental feature.

A competition-centered CP3 should teach, not merely demand, the following:
\begin{itemize}
    \item rapid problem-set scanning and triage;
    \item estimation of expected solve time and implementation risk;
    \item strategic switching and stopping rules;
    \item role allocation, handoffs, and one-computer team coordination;
    \item concise explanation of proof ideas and edge cases;
    \item management of penalties, uncertainty, and emotional recovery after failed submissions;
    \item post-contest reconstruction of decisions, not only editorial review of solutions.
\end{itemize}

Contest strategy literature offers concrete practices for this stage \cite{trotman2008strategy,bloomfield2016guide}. However, course grades should generally be criterion-referenced rather than rank-referenced. A student can demonstrate improved triage, communication, and solution quality even when placed against unusually strong peers. The competition is the authentic activity; peer rank need not be the grading construct.

\section{Assessment Architecture}
\label{sec:assessment}

The assessment architecture uses multiple forms of evidence because no single measure captures the six outcome clusters. The online judge provides high-frequency evidence of functional correctness and efficiency. Breadth requirements prevent students from earning a passing grade through narrow specialization. Oral interviews probe understanding and authorship. Optional clean-code submissions make readability visible. In-class pair problems and reflections provide formative evidence about reasoning. Timed contests are reserved for later or low-stakes use.

\subsection{Mastery-Oriented Online-Judge Problem Sets}

CP1 uses twelve weekly problem sets with four problems each, for 48 possible accepted solutions \cite{dickey2026syllabus}. Each problem is scored dichotomously: the submission either satisfies the hidden correctness, time, and memory tests or it does not. Students may resubmit until the deadline, and unsuccessful attempts do not directly reduce the grade.

This structure has several strengths. It preserves a meaningful performance standard: a partially correct or asymptotically infeasible program does not receive credit. It provides immediate feedback at a scale that hand grading cannot match \cite{alamutka2005automatic,wasik2018taxonomy}. It also supports mastery by separating the number of failed attempts from the eventual evidence of competence \cite{bloom1968mastery,kulik1990mastery}.

The same structure has important limitations. Verdict-driven work can become unprincipled trial and error. Hidden tests can obscure whether the defect is conceptual or local. Binary scoring cannot distinguish an elegant proof-backed solution from copied code or a brittle implementation that happens to pass. Moreover, the count of solved problems is a measure of accumulated successful performances, not a calibrated interval measure of proficiency. Our design therefore treats judge acceptance as necessary evidence for implementation outcomes, but insufficient evidence for the full construct.

Problem selection is part of the assessment design. A weekly set should form a deliberate progression rather than a random sample:
\begin{enumerate}
    \item an entry problem that rehearses the canonical structure;
    \item a second problem with a changed representation or constraint;
    \item a transfer problem with misleading surface context or a required combination;
    \item an optional stretch problem that supports advanced learners without becoming a hidden prerequisite for passing.
\end{enumerate}

Difficulty labels from platforms can inform selection, but instructor analysis is required because nominal difficulty does not reveal which prerequisite, insight, or implementation hazard makes a problem difficult for a particular cohort.

\subsection{A Breadth Gate Against Strategic Narrowing}

Point accumulation alone invites rational specialization. A student might repeatedly solve familiar categories, avoid dynamic programming or graphs, and still reach a numerical threshold. The grading architecture therefore requires at least one accepted problem from 11 of the 12 distinct problem sets for a passing grade, in addition to the total-point requirement \cite{dickey2026syllabus}. An earlier version used a similar ten-topic gate \cite{dickey2022cp1proposal}.

This rule is a simple but consequential alignment device. The course claims that classification across a broad survey is enduring; the grading system must therefore require evidence across the survey. The gate should be communicated as a coverage standard rather than a punitive technicality, and the learning-management system should make progress visible. An even stronger future version would connect each problem to outcome tags and require minimum evidence across both content families and process outcomes.

\subsection{Sampled Oral Code Interviews}\label{sec:assessment:interviews}

Accepted code establishes functional performance, but it does not establish that the student can reconstruct the reasoning or that the submitted work reflects the student's own understanding. CP1 therefore samples approximately ten students each week (10\%) for a 5-15 minute interview over one solution accepted during the preceding week. Selection is random subject to the requirement that every student be interviewed at least once, and the student explains the solution without viewing the submitted code \cite{dickey2026syllabus}. The interviews are conducted in instructor or head-TA office hours, so the recurring workload is approximately 1.7 assessor-hours per week for a cohort near 100.

The interview uses four core prompts and one optional followup:
\begin{enumerate}
    \item What is the problem asking you to compute? Explain a sample input and output.
    \item What is a straightforward or brute-force approach?
    \item Explain the accepted approach, including the decisive insight and the relevant CP paradigm or terminology.
    \item What are its time and space bounds, and why are they sufficient for the stated constraints?
    \item (optionally) What is one thing you got wrong initially that you later corrected? (This prompt can distinguish reconstructable process knowledge from a brittle surface-level account and can provide additional context when authorship is uncertain.)
\end{enumerate}
Neutral follow-ups such as ``Why?'', ``What happens next?'', or ``Which assumption are you using?'' probe an unclear response without turning the interview into a second programming assignment. In repeated use, the instructor has found this short sequence effective at distinguishing students who can reconstruct their work from students whose understanding is too shallow to support even a focused follow-up. That observation motivates the practice, although it is not presented here as measured evidence of effectiveness.

When an interview does not produce sufficient evidence of understanding, the sampled solution loses credit because judge acceptance no longer provides adequate evidence for that assignment. The submission is also reviewed for possible plagiarism or unauthorized assistance. This second step is an investigation, not a finding: staff inspect the code, similarity evidence, submission history, and other relevant information under the institution's ordinary academic-integrity process. A weak interview by itself does not establish misconduct.

The protocol can be strengthened without expanding it into a full oral examination. We propose publishing a compact three-dimensional rubric (\Cref{tab:oralrubric}), calibrating interviewers on several scripted cases, and permitting a brief targeted probe when the initial response is ambiguous (especially when interviewed by staff other than the instructor). A short follow-up interview is appropriate when staff cannot distinguish limited understanding from a communication or prompt-alignment problem. \Cref{app:interview} gives an implementation-ready version. Oral assessment research supports this attention to prompt structure and scoring criteria \cite{joughin1998oral}, while layered assessment addresses the security limits of submitted code in an environment with public solutions and generative tools \cite{dawson2021security}, \cite{dickey2025plagiarism}.

\subsection{Clean Code, Explanation, and Reflection}

CP rewards concise implementation, but competition code can be opaque. The course offers modest bonus credit for solutions that are cleaned, named clearly, and commented so that another computing student can follow the organization \cite{dickey2026syllabus}. This component helps separate ``code that wins a contest'' from ``code that communicates.'' It also creates a bridge to software-engineering values without pretending that a short algorithmic program exercises full software-engineering practice.

Because comments can be produced mechanically, the educational target is explanatory coherence rather than comment density. A compact rubric can ask whether the code's decomposition matches the algorithm, names carry domain meaning, invariants or non-obvious steps are explained, dead debugging code is removed, and the declared complexity matches the implementation. Selected solutions can be compared anonymously in class to show that multiple correct implementations embody different tradeoffs.

Metacognitive prompts are similarly short but consequential: What clue first suggested the eventual paradigm? Which false start was most informative? What would change if the largest input were ten times larger? What test case distinguishes your solution from the nearest incorrect approach? Reflection should be attached selectively to high-value problems so that it remains substantive rather than becoming repetitive compliance \cite{flavell1979metacognition,zimmerman2002srl}.

\subsection{A Bounded Generative AI Intervention}

The course includes a designated week in which students may use generative AI if they document chat links and reflect on how the tool contributed, failed, or changed their process \cite{dickey2026syllabus}. This is preferable to an incoherent policy in which students are told only to avoid tools that are widely available and increasingly capable. AI code generation has already complicated the relationship between submitted code and student understanding \cite{finnieansley2022robots,becker2023programming,lau2023adapt}.

We treat the intervention as a bounded instructional experiment with explicit guardrails. A well-designed week can ask students to perform some subset of the following, potentially under supervision:
\begin{enumerate}
    \item predict the algorithm before prompting the model;
    \item interrogate a model-generated explanation or code sample;
    \item construct a counterexample to an incorrect claim;
    \item compare model and student complexity analyses;
    \item revise prompts and document how verification changed confidence;
    \item complete an oral explanation without model access.
\end{enumerate}

The AI-Lab intervention literature provides a relevant scaffold for guided, reflective use \cite{dickey2025ailab}. Claims about the CP-specific implementation remain questions for direct evaluation. Chat-link requirements also raise privacy, persistence, and platform-access issues; an export or instructor-provided log format is safer than requiring public links. The pedagogical objective here is verification literacy and self-regulation, not surveillance.

\subsection{The Role of Timed Assessment}

Timed work is not absent from the sequence. CP1 may use a short, low-stakes bonus contest or diagnostic near the end of the term, CP2 may use mixed-topic timed sets, and CP3 should use repeated authentic simulations. The crucial distinction is whether time pressure is an instructional target and whether the score is interpreted accordingly.

A CP2 timed diagnostic can answer, ``Which schemas are slow to retrieve?'' It should be followed by a retrospective and should not dominate the course grade. A CP3 contest can answer, ``Can this student or team triage and execute under the target conditions?'' Its rubric may include accepted problems, time to first acceptance, unnecessary submissions, strategic switches, communication, and the quality of post-contest analysis. Raw rank is useful contextual information but should not be the sole grade because it is relative to the field rather than an absolute learning standard.

\section{Pedagogical Architecture}
\label{sec:pedagogy}

\subsection{A Three-Part Class Meeting}

Each week follows a recurring three-part lesson pattern \cite{dickey2025notes}. In the two-meeting format, roughly 15--20 minutes introduce the week's conceptual structure; most remaining time is devoted to interactive analysis of problems; and approximately 25 minutes near the end of the second meeting are used for paired problem solving. The exact timings are adjustable, but the functional sequence is important.

\textbf{1) Compact conceptual exposition.} The instructor names the paradigm or structure, specifies the conditions that make it applicable, contrasts it with nearby alternatives, and models a systematic analysis. For greedy algorithms, for example, the mini-lesson should not stop at ``make the locally best choice.'' It should distinguish candidate orderings, candidate heuristics, exchange arguments, and failure cases. The purpose of this part is to supply a vocabulary and an initial schema, not to pre-solve every possible homework variation.

\textbf{2) Think-pair-share problem analysis.} Students first inspect a problem independently, then compare models and candidate approaches with a partner, and finally contribute to a whole-class discussion. Peer discussion can improve answers even when neither partner initially has the correct response because the interaction externalizes reasoning and creates opportunities to resolve contradictions \cite{smith2009peer}. In CP, the share phase should require more than naming an algorithm. Students, as a collective, must point to a constraint or structural clue, state the brute-force baseline, explain why it fails or succeeds, and defend the proposed improvement in response to peers and instructor questions.

Incorrect approaches are valuable when handled with psychological safety. A wrong greedy rule, for example, can reveal the missing exchange property more clearly than a correct answer offered without justification. The instructor should normalize peer interaction, group reasoning, and revision, separate ideas from identities, and ask the class to produce the smallest counterexample, once a proposed approach is recognized as flawed. This converts public error from a status event into shared diagnostic evidence.

\textbf{3) Paired whole-problem solving.} Students receive an approachable problem and a randomly or deliberately assigned partner (with random pairing emphasized in CP1 and deliberate team assignment used in CP3 to address team strengths and weaknesses as a learning objective). They must move from statement to model, algorithm, complexity, and implementation while instructors and teaching assistants circulate. This is a ``junior version of the whole game'' \cite{perkins2009whole}: the task remains authentic, but its difficulty and support are calibrated. Pair work is not merely social; it forces learners to articulate decisions that might otherwise remain tacit.

The three parts correspond to increasing ICAP engagement: exposition can be active if students predict and annotate; problem analysis is constructive; peer comparison and joint implementation are interactive \cite{chi2009active,chiwylie2014icap}. The class should spend the minority of time on uninterrupted lecture and the majority on producing, testing, and revising explanations.

\subsection{Fading Scaffolds Across the Semester}

Early in CP1, the instructor can provide a structured worksheet:
\begin{enumerate}
    \item restate the output in one sentence;
    \item work one small example by hand;
    \item list the maximum input sizes;
    \item write the brute-force algorithm and complexity;
    \item identify repeated work or exploitable structure;
    \item propose a representation and efficient algorithm;
    \item state the correctness idea or invariant;
    \item list boundary tests.
\end{enumerate}

Over time, prompts are removed, problem topics are mixed, and students take responsibility for challenging one another's assumptions. This fading avoids two extremes: discovery with insufficient guidance and recipe following without transfer \cite{wood1976scaffolding,collins1989apprenticeship}. CP2 can preserve only a compact checklist, while CP3 expects teams to execute the routine rapidly and implicitly.

Contrasting cases are especially useful. Present two problems with nearly identical stories but different constraints, or two algorithms that both pass small examples but diverge asymptotically. Ask students to identify the minimum change that flips the correct paradigm. Such comparisons train attention to deep structure and support future learning \cite{bransford1999transfer,streveler2008conceptual}.

\subsection{Office Hours as Coached Problem Solving}

Office hours are part of the mastery system rather than a last-resort debugging service. Staff are expected to move students toward a correct, explainable solution without taking over the student's reasoning \cite{dickey2025notes}. A useful coaching progression is:
\begin{enumerate}
    \item \textbf{Diagnose:} ask the student to state the problem, current model, and evidence from failed tests;
    \item \textbf{Elicit:} ask for the brute-force method, constraints, and suspected bottleneck;
    \item \textbf{Focus:} direct attention to one overlooked property or construct a counterexample;
    \item \textbf{Scaffold:} offer a representation, subgoal, or analogous miniature problem rather than a code fragment;
    \item \textbf{Fade:} return control and require the student to complete the reasoning;
    \item \textbf{Verify:} ask for complexity or, if time, an invariant or discriminating test before implementation resumes.
\end{enumerate}

This approach resembles cognitive apprenticeship: make expert processes visible, coach performance, and gradually withdraw support \cite{collins1989apprenticeship}. Staff training matters. Without shared norms, one teaching assistant may reveal a solution immediately while another withholds even a productive question. \Cref{app:ta} provides a compact protocol.

\subsection{Participation, Talk Distribution, and Inclusion}

Interactive learning can reproduce status hierarchies when experienced or fast-speaking students consistently fill the available conversational space. We address this problem with a low-overhead diagnostic rather than attempting to track roles and turn-taking throughout a 50--100 student course. The activity was developed during the original CAP redesign in consultation with an engineering-education faculty member as a practical response to a participation pattern that is otherwise difficult to diagnose at this scale.

During the opening portion of the semester, students complete a four-person ``Make a Sandwich'' assignment worth the same number of points as one programming problem. The group is asked to ``create an algorithm'' for assembling an intentionally absurd sandwich from peanut butter and jelly, ham and an American-cheese single, Nutella, Funyuns, and whole-grain bread in a specified order. The prompt deliberately leaves \emph{algorithm} undefined. Students must negotiate what level of precision, ordering, and completeness is required, much as they must surface and reconcile ideas during think-pair-share. The unusual ingredients make the task memorable and visibly low in disciplinary stakes, while grading it for correctness and completeness gives even skeptical students a reason to engage seriously.

The discussion is recorded with students' knowledge, and the automated meeting-transcription tool Fireflies.ai supplies a coarse estimate of each speaker's share of the conversation. Speaking time is not part of the grade. The instructor uses the estimate together with direct observation to distinguish students who tend to occupy conversational space from those who rarely enter it. This secondary seating purpose is not disclosed to students because disclosing it would change the interaction being sampled; it is used only for grouping, never for public labels, research data, or participation penalties. The recording and transcript are accessible only to the instructor and are deleted after the seating assignment has been made.

At the beginning of week two, students receive paper name tents and are assigned to the left or right half of the classroom for the semester, with seating otherwise unrestricted. The assignment separates the highest-airtime students from those least likely to speak, so that quieter students are not continually paired with someone who immediately fills every silence. The instructor also preserves individual think time and occasionally calls on a randomly selected group rather than relying only on volunteers. In repeated offerings, the instructor has observed a characteristic progression: one side contributes very little during the first think-pair-share, while by the end of the term both sides participate at comparable levels. This is a practitioner observation rather than a quantified outcome, but it explains why the mechanism has been retained.

The method is intentionally coarse. Speaking time does not measure idea quality, confidence, language fluency, or learning, and automated diarization can be wrong. Those limitations are controlled by using the output only for a one-time half-room assignment, combining it with instructor judgment, avoiding individual participation grades or public classification, and restricting access to the recording and transcript. More elaborate systems of rotating roles, participation tokens, or continuous talk tracking can be useful in smaller courses, but their administrative cost is disproportionate here. The retained design uses one brief diagnostic to create more balanced opportunities for the interactive reasoning supported by cooperative-learning research \cite{johnson1998cooperative,springer1999smallgroup}.

\subsection{Attendance and Access to Learning Materials}

Attendance is voluntary and is not part of the grade. This choice fits an interactive course: students who do not intend to participate are not required to occupy the room, while those who attend are expected to analyze problems, compare reasoning, and work with peers. At the same time, the course creates a modest positive incentive for attendance by releasing each session's slides and recording through an easy end-of-class learning-management-system item. The consequence of an ordinary absence is limited to losing automatic access to that session's supplementary materials; assignments, deadlines, and announcements are in the course system, and grade opportunities remain available \cite{dickey2025notes},\cite{dickey2026syllabus}.

The policy reflects the instructional role of the meetings and repeated local experience that unrestricted recordings can become a substitute for attendance even when the central value of class is live peer reasoning and immediate coaching. The broader literature supports both parts of this concern: students use recordings productively for review and assessment preparation, while availability can reduce live attendance; computing-education studies also find distinct patterns of live and video participation rather than a single universal use case \cite{morris2019recordings,ihantola2020preferences,picardo2021recordings}. The course therefore treats recordings as contingent review resources instead of assuming that universal release is pedagogically neutral.

The incentive is paired with a clear exception process. Students who miss class for a university-recognized or otherwise legitimate reason are explicitly told to contact the instructor and can receive the materials. Required accommodations are honored, and students who come to office hours seeking to recover missed learning are generally given the relevant resources. In a large-enrollment course, the process relies on established absence categories, accommodation procedures, and reasonable instructor judgment rather than an open-ended parallel attendance system. It also asks adult students to communicate when they need an exception. In this implementation, the arrangement preserves voluntary attendance while keeping the live, interactive meeting as the easiest route to the full set of course resources.

\section{Constructive Alignment and Theory of Action}
\label{sec:alignment}

\begin{figure*}[t]
    \centering
    \resizebox{0.97\textwidth}{!}{%
    \begin{tikzpicture}[
        node distance=7mm and 8mm,
        box/.style={draw, rounded corners=2pt, align=center, minimum height=9mm, text width=28mm, fill=gray!8},
        smallbox/.style={draw, rounded corners=2pt, align=center, minimum height=8mm, text width=24mm, fill=gray!4},
        arrow/.style={-{Latex[length=2.2mm]}, thick},
        dashedarrow/.style={-{Latex[length=2.2mm]}, thick, dashed}
    ]
        \node[box] (model) {Conceptual model\\and worked contrast};
        \node[box, right=of model] (analyze) {Individual and\\peer problem analysis};
        \node[box, right=of analyze] (implement) {Whole-problem\\implementation};
        \node[box, right=of implement] (judge) {Online-judge\\feedback and retry};
        \node[box, right=of judge] (explain) {Oral explanation,\\tests, and reflection};

        \draw[arrow] (model) -- (analyze);
        \draw[arrow] (analyze) -- (implement);
        \draw[arrow] (implement) -- (judge);
        \draw[arrow] (judge) -- (explain);
        \coordinate (loopright) at ($(explain.north)+(0,8mm)$);
        \coordinate (loopleft) at ($(model.north)+(0,8mm)$);
        \draw[arrow, rounded corners=2mm] (explain.north) -- (loopright) -- node[midway,above,align=center]{distributed practice and fading scaffolds} (loopleft) -- (model.north);

        \node[smallbox, below=13mm of analyze] (cp1) {CP1: broad schemas,\\high-to-moderate support};
        \node[smallbox, right=10mm of cp1] (cp2) {CP2: integration,\\ambiguity, transfer};
        \node[smallbox, right=10mm of cp2] (cp3) {CP3: speed, triage,\\team communication};
        \draw[arrow] (cp1) -- (cp2);
        \draw[arrow] (cp2) -- (cp3);

        \node[draw, rounded corners=2pt, fit=(model)(analyze)(implement)(judge)(explain), inner sep=4mm] (layerone) {};
        \node[font=\bfseries, anchor=west] at ($(layerone.north west)+(0,12mm)$) {Layer 1: algorithmic problem-solving competence};
        \node[draw, rounded corners=2pt, fit=(cp3), inner sep=3mm, label={[align=center]below:\textbf{Layer 2 added explicitly}}] {};
    \end{tikzpicture}}
    \caption{Theory of action for deferred competition. CP1 and CP2 repeatedly enact the full algorithmic cycle with increasing independence; CP3 adds the contest-performance conditions as an explicit target rather than an implicit prerequisite.}
    \label{fig:theory}
\end{figure*}
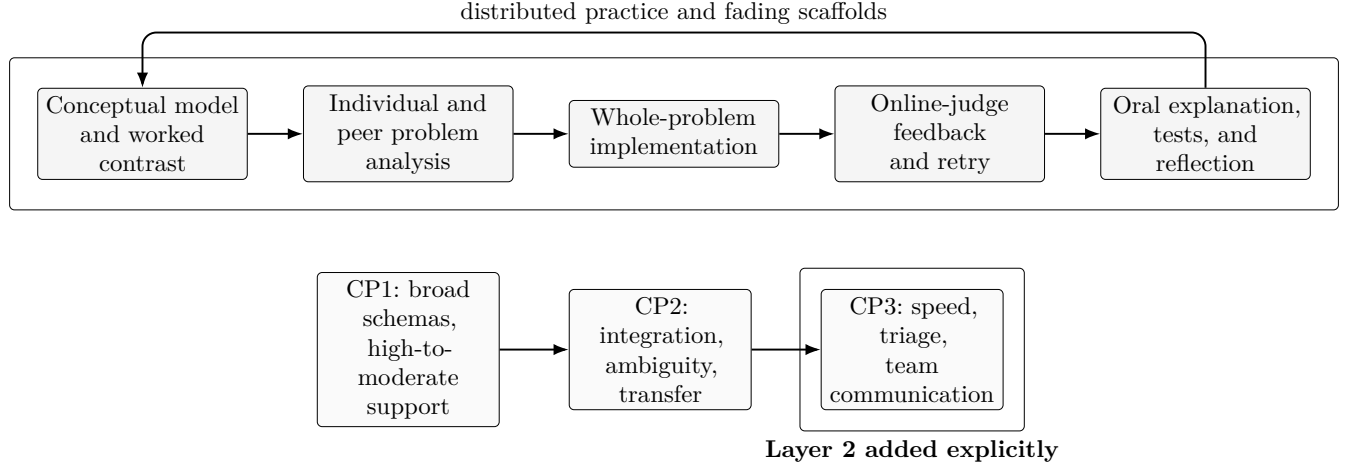

\Cref{fig:theory} summarizes the curriculum's theory of action. A weekly mastery cycle connects a compact conceptual model to collaborative analysis, whole-problem practice, judge feedback, human explanation, and reflection. Across CP1 and CP2, scaffolds fade and problems become more integrated. CP3 adds a contest-performance layer only after students have repeatedly enacted the underlying algorithmic cycle.

\Cref{tab:alignment} makes the alignment claim auditable. Every consolidated outcome is practiced in class and assessed by at least one direct source of evidence. Conversely, each major graded component has an identified outcome function. This avoids a common failure in which collaborative reasoning is praised in the syllabus but invisible in assessment, or accepted code is treated as evidence for metacognition that it cannot provide.

\begin{table*}[t]
    \centering
    \caption{Constructive alignment matrix for the consolidated CP1 outcomes. Filled circles indicate primary evidence; open circles indicate secondary or formative evidence.}
    \label{tab:alignment}
    \scriptsize
    \begin{tabularx}{\textwidth}{@{}P{0.20\textwidth} *{6}{>{\centering\arraybackslash}p{0.073\textwidth}} Y@{}}
        \toprule
        \textbf{Learning experience or assessment} & \textbf{O1 Classify} & \textbf{O2 Design} & \textbf{O3 Analyze} & \textbf{O4 Implement} & \textbf{O5 Debug} & \makecell{\textbf{O6}\\\textbf{Explain}\\\textbf{transfer}} & \textbf{Design function} \\
        \midrule
        Mini-lessons and contrasting worked cases & $\circ$ & $\circ$ & $\circ$ &  &  & $\circ$ & Make expert cues and reasoning structures visible \\
        Think-pair-share analysis & $\bullet$ & $\bullet$ & $\circ$ &  &  & $\bullet$ & Construct and defend models before coding \\
        In-class paired whole problem & $\circ$ & $\bullet$ & $\bullet$ & $\bullet$ & $\circ$ & $\circ$ & Practice the complete cycle with immediate coaching \\
        Weekly online-judge problems & $\circ$ & $\bullet$ & $\bullet$ & $\bullet$ & $\bullet$ &  & Require executable correctness and efficiency across distributed practice \\
        Breadth gate & $\bullet$ & $\circ$ & $\circ$ & $\bullet$ & $\circ$ &  & Prevent narrow topic specialization from satisfying course mastery \\
        Sampled oral interview & $\bullet$ & $\bullet$ & $\bullet$ & $\circ$ & $\circ$ & $\bullet$ & Elicit reasoning, complexity, traceability, and authorship evidence \\
        Custom-test/debugging prompt &  & $\circ$ & $\circ$ & $\circ$ & $\bullet$ & $\circ$ & Turn judge failure into specification-based diagnosis \\
        Clean-code submission/code comparison &  & $\circ$ & $\circ$ & $\bullet$ & $\circ$ & $\bullet$ & Make representation and communication quality visible \\
        Metacognitive reflection & $\circ$ & $\circ$ & $\circ$ &  & $\circ$ & $\bullet$ & Abstract cues and monitor strategy for transfer \\
        Low-stakes timed diagnostic & $\circ$ & $\circ$ & $\circ$ & $\circ$ & $\circ$ & $\circ$ & Diagnose retrieval latency without making speed the introductory construct \\
        \bottomrule
    \end{tabularx}
\end{table*}

The alignment also clarifies the purpose of course policies. Repeated submission is more than a convenience; it supports implementation mastery. The breadth gate is not an arbitrary hurdle; it operationalizes a survey-course outcome. Pair problem solving is not an entertainment break; it supplies interactive whole-task practice. Oral interviews supply direct evidence of explanation and reasoning while also identifying submissions that warrant further authorship review. CP3 contests are not ceremonial finals; they assess a newly taught performance layer.

\section{Implementation Evolution and Design Tradeoffs} \label{sec:evolution}
The curriculum changed substantially after the initial CAP design. The 2022 version centered on weekly accepted problems and explicit alignment of broad outcomes \cite{dickey2022design}. The course proposals then translated that design into official CP1, CP2, and CP3 offerings, including a breadth requirement, clean-code opportunities, and a distinct contest-preparation capstone \cite{dickey2022cp1proposal},\cite{dickey2022cp2proposal},\cite{dickey2023cp3proposal}. By Spring 2026, CP1 used 48 possible problems, sampled oral interviews, custom-test debugging as an objective, a bounded GenAI week, and the current two-meeting pedagogical pattern \cite{dickey2026syllabus}.

This history establishes operational feasibility under the local conditions: the architecture has been implemented across nine offerings, institutionalized as a three-course sequence, and revised in response to new needs such as authorship verification and generative AI. It also shows that course design is not completed when a proposal is approved. The official proposals made the sequence permanent, while the redesigned syllabus replaced the earlier temporary-course materials and continued to evolve through implementation.

\Cref{tab:tradeoffs} records the function of each major feature, the boundary condition that deserves attention, and a response proportionate to the scale of the problem. The same feature may reasonably be implemented differently in another institution, but the educational function should remain explicit.

\begin{table*}[t]
\centering
\caption{Design tradeoffs, boundary conditions, and proportionate implementation responses.}
\label{tab:tradeoffs}
\footnotesize
\begin{tabularx}{\textwidth}{P{0.18\textwidth} Y Y Y}
\toprule
\textbf{Design feature} & \textbf{Educational function} & \textbf{Boundary condition} & \textbf{Proportionate response or evaluation} \\
\midrule
Binary accepted-problem grading & Enforces correctness and efficiency at scale; retry supports eventual mastery & Passing verdicts do not reveal reasoning or distinguish systematic debugging from trial and error & Retain judge acceptance as implementation evidence; add oral sampling, occasional test-generation prompts, and problem/outcome tags \\
Breadth gate & Requires engagement across the survey rather than strategic specialization & One easy solve is a thin sample of a topic; students can overlook late coverage requirements & Display progress continuously and use richer evidence for selected core families \\
Sampled oral interviews & Supplies understanding evidence and a bounded trigger for authorship review & Brief retrieval can be noisy; interviewer prompts and schedules vary & Publish a compact rubric, calibrate interviewers, use one targeted follow-up when evidence is ambiguous, and treat integrity review as investigation \\
Clean-code bonus & Makes readability and explanation visible & Optional status can make communication appear peripheral; comments can be superficial or generated & Score explanatory coherence rather than comment quantity; consider one required sample in future versions \\
Voluntary attendance & Avoids grading physical presence and protects the interactive climate from disengaged attendance & Students can miss high-value live practice or in-class announcements & Keep essential course information universally available and make the live meeting visibly valuable \\
Attendance-linked material release & Creates a modest incentive for live participation without an attendance grade & Legitimate absences and accommodations require access; recordings can support review while also substituting for attendance & Release automatically to attendees and provide materials through a clearly stated exception and help-seeking process \\
Speaking-time diagnostic and half-room seating & Redistributes conversational opportunity with one low-overhead intervention & Talk time is a coarse proxy; automated diarization can err; recordings require careful handling & Use only for coarse grouping, never for participation grades or public labels; combine with observation and restrict access to recordings \\
High-support office hours & Keeps struggle productive and reduces abandonment & Staff can provide inconsistent levels of help or solve too much of the task & Train staff on the diagnose-elicit-focus-scaffold-fade-verify protocol and calibrate with short cases \\
Bounded GenAI week & Develops verification literacy and reflective tool use & Model variability, access, privacy, and substitution for thinking can affect the activity & Use common guardrails, reflection, and oral verification; evaluate the intervention separately from the surrounding course \\
Deferred competition & Protects foundational practice and the validity of introductory assessment & Advanced entrants can be under-challenged; speed does not emerge automatically & Offer challenge problems or accelerated entry and explicitly teach contest performance in CP3 \\
\bottomrule
\end{tabularx}
\end{table*}

\subsection{Implementation Evidence and Open Questions}

Think-pair-share, paired whole-problem solving, high-support office hours, sampled interviews, and the participation-balancing diagnostic have been retained because they repeatedly fit the instructional goals and operating constraints. The current enrollments also demonstrate that online-judge grading and distributed teaching-assistant support can sustain the architecture at the scale described in \Cref{sec:method}. These observations are useful evidence about enactment: the features can be scheduled, staffed, and integrated into an official sequence.

They do not yet quantify changes in student learning or establish that this sequence outperforms a contest-first alternative. The most appropriate next step is therefore a focused evaluation that reuses existing course activities rather than an attempt to instrument every possible outcome at once.

\section{Evaluation Agenda} \label{sec:evaluation}

The central design claim can be studied at several levels of effort. A full longitudinal mixed-methods study would be valuable, but it is not the only meaningful next step and would impose substantial collection, scoring, and governance costs. \Cref{tab:minstudy} therefore presents a set of nested study designs that reuse existing assignments, interviews, and contest logs. An instructor can begin with one minimum-viable design and add the extended measures in \Cref{app:evaluation} only when the intended claim requires them.

\subsection{Research Questions and Testable Conjectures}

Four research questions organize the program of work.
\begin{itemize}
    \item \textbf{RQ1 -- Algorithmic learning.} To what extent do students develop the ability to classify, decompose, analyze, implement, debug, and explain novel CP problems across CP1 and CP2?
    \item \textbf{RQ2 -- Contest transfer.} After Layer 1 instruction, what additional changes occur when CP3 explicitly teaches contest performance? Examples include faster scanning of a problem set, earlier selection of solvable problems, less time committed to eventually abandoned problems, improved time to first acceptance and total solves, clearer team handoffs, and better recovery after a failed submission, while untimed explanation and complexity reasoning remain stable.
    \item \textbf{RQ3 -- Participation and motivation.} How do the sequence and its classroom climate relate to self-efficacy, belonging, persistence, and participation for students with different levels of prior CP experience?
    \item \textbf{RQ4 -- Feasibility.} What staffing, instructional time, calibration, and technological infrastructure are required to enact the aligned design at the intended scale?
\end{itemize}

The theory of action yields four corresponding conjectures. (1) Improvement should appear on unfamiliar, untimed transfer tasks rather than only on rehearsed problem families. (2) Oral reasoning should contribute information beyond accepted-problem count because the two assessments sample different parts of the construct. (3) CP3 should improve repeated contest indicators and team strategy without reducing untimed reasoning quality. (4) Finally, the mastery-oriented entry should narrow early participation differences between novices and experienced competitors. These are testable design conjectures rather than conclusions from the present artifact analysis.

\subsection{Minimum-Viable Study Designs}

\begin{table*}[t]
    \centering
    \caption{Nested evaluation designs, ordered from lowest to highest additional burden.}
    \label{tab:minstudy}
    \scriptsize
    \begin{tabularx}{\textwidth}{P{0.10\textwidth} Y Y P{0.23\textwidth}}
        \toprule
        \textbf{Design} & \textbf{Additional data collection} & \textbf{Primary inference} & \textbf{Burden and main limitation} \\
        \midrule
        End-of-CP1 evidence audit & One unfamiliar 20--30 minute analysis task in the final week (e.g., classify; give brute-force and efficient approaches; justify complexity; provide one edge case), scored with a compact rubric and compared with existing solve counts, breadth completion, and sampled interview scores & Whether judge performance, topic breadth, and articulated reasoning converge at the end of the course; identifies outcomes weakly represented by the current grading system & Low; one task and one scoring pass. It describes end-point alignment but cannot estimate growth \\

        CP1 pre/post transfer study & Parallel versions of that task in the first and final weeks, plus one prior-experience item; reuse the rubric & Within-student change in classification, design, complexity, and verification, and the relation of that change to ordinary course evidence & Moderate; two short administrations and rubric scoring. Repeated-task and maturation effects remain \\

        CP3 contest-transfer study & Use logs from the first two and last two course contests, a one-page team retrospective after each (e.g., initial problem order, switches, and handoffs), and one untimed mixed-technique anchor task before and after CP3 & Whether explicit contest instruction changes triage, time allocation, handoffs, penalties, time to first acceptance, and solves while preserving untimed reasoning & Low-to-moderate because contests already occur. Contest difficulty and self-selection into CP3 must be modeled or described \\

        Full sequence study & Follow consenting students through CP1--CP3 with common transfer tasks, repeated contests, motivation measures/brief surveys, selected interviews, and implementation-cost logs & The complete theory of action, including Layer 1 growth, Layer 2 transfer, participation, motivation, and sustainability & High; multi-semester participation, attrition, calibration, and data governance make this a separate research project \\
        \bottomrule
    \end{tabularx}
\end{table*}

For the next offering, the CP1 pre/post transfer study is the strongest minimum-viable choice because it addresses the principal introductory-course claim with only two common tasks. Pairing it later with the CP3 contest-transfer study would test the full sequencing argument without requiring the full longitudinal design. The end-of-course audit remains useful when even pretesting is infeasible.

\subsection{Core Measures of Layer 1 Competence}

A common novel task should require students to restate the problem, give a brute-force baseline, identify structural cues, design an efficient solution, connect complexity to the constraints, and provide at least one discriminating test. One task can be code-producing and another can be analysis-only, but a minimum-viable study may use a single carefully designed prompt scored on the same compact dimensions as the oral interview. Parallel pre/post tasks should share deep structure and expected difficulty without differing only in names or story context \cite{bransford1999transfer,schoenfeld1985problem}.

Existing course evidence then becomes analytically useful rather than being discarded. Accepted problems measure accumulated implementation performance, the breadth gate measures distribution across topics, and interviews measure reconstructable understanding. Their convergence or disagreement is itself informative. Online-judge traces can be added selectively when the research question concerns debugging; a high submission count should be interpreted alongside code changes or custom tests because it can represent either productive iteration or unstructured resubmission \cite{alamutka2005automatic,keuning2019feedback,hattie2007feedback,shute2008feedback}.

\subsection{Core Measures of Layer 2 Competence}

CP3 already produces most of the required evidence. Across repeated contests, automatically available indicators include solves, penalties, time to first acceptance, submission sequences, and problem order. A short team retrospective can record initial triage decisions, why the team persisted or switched, how ideas were handed off, and where avoidable idle time occurred. These measures answer the concrete forms of RQ2 more directly than leaderboard rank alone.

At least two early and two late contests should be compared because a single event is sensitive to problem-set composition and day-level variation. A brief untimed anchor task before and after CP3 tests whether faster performance coexists with stable explanation, complexity, and debugging practices. Performance-consistency research likewise supports repeated events and absolute indicators rather than interpreting one rank as a stable trait \cite{luodickey2026consistency}.

\subsection{Extended Evaluation Modules}\label{sec:evaluation:extended}

\Cref{app:evaluation} contains the full evidence matrix and optional modules for motivation and participation, implementation fidelity and cost, and research data governance. These modules are important when the study makes claims about access, transferability, or mechanisms, but they need not be collected merely to conduct the minimum-viable Layer 1 or Layer 2 studies above.

\section{Limitations and Boundary Conditions} \label{sec:limitations}

This paper is a design case assembled from curriculum artifacts and practitioner reflection. It does not contain student-level outcome data, a comparison group, or an independent fidelity study. The multi-year implementation history supports plausibility and operational feasibility, while claims about learning, motivation, participation, or contest transfer remain to be tested through the designs in \Cref{sec:evaluation}.

The case also reflects one institutional context. Students enter after \prerequisitecourse{}, the sequence is supported by an online judge and a sizable teaching-assistant team, and each course is a compressed two-credit offering. A program serving novice programmers, operating with fewer staff, or lacking automated judging may need a narrower scope and a different assessment sample. The transferable object from this paper is the alignment among outcomes, evidence, pedagogy, and the sequencing of competition, rather than an invariant list of weekly topics.

The distinction between algorithmic and contest-performance competence is analytical rather than absolute. Speed develops through ordinary practice, communication improves during collaborative CP1 activities, and contests can reveal gaps in algorithmic schemas. The model claims a difference in instructional emphasis and measurement. Optional low-stakes contests can therefore coexist with deferred consequential competition, especially for experienced students who need additional challenge.

The breadth of CP also limits what any survey course can claim. A few accepted tasks cannot represent every algorithm family, problem tags are contest-community interpretations, and many strong tasks admit multiple approaches. The breadth gate is best understood as evidence that students engaged across the designed survey, not as proof of comprehensive algorithmic expertise. Oral reasoning, mixed-topic tasks, and transfer measures improve this evidence without making it exhaustive.

Several locally retained mechanisms depend on clearly stated boundaries. Attendance-linked material release has provided a practical incentive for an interactive course without grading attendance, while its exception process must continue to support recognized absences and accommodations. The sandwich diagnostic has given the instructor a low-overhead way to redistribute discussion opportunities, while speaking time remains a coarse grouping signal rather than a measure of learning or a student trait. Sampled interviews provide direct evidence of understanding and a useful trigger for examining a submission, while the resulting plagiarism review remains an investigation rather than a presumption of misconduct. These qualifications preserve the direct benefits that motivated the practices and identify the conditions under which they should be adapted elsewhere.

Finally, generative AI continues to change the relationship among authorship, implementation, explanation, and feedback. A bounded intervention is a useful design probe, but model capabilities, access conditions, and institutional rules change rapidly. The durable outcomes are the capacity to specify a problem, inspect an algorithm, test generated code, explain its assumptions, and remain responsible for submitted work \cite{finnieansley2022robots,becker2023programming,lau2023adapt}, \cite{dickey2025ailab}.

\section{Conclusion} \label{sec:conclusion}
Competitive programming education need not choose between authentic contest culture and rigorous course design. It can make a more precise choice about sequence. CP tasks provide unusually rich, objective, and scalable opportunities to practice the complete path from problem statement to executable solution. Timed contests add another valuable layer: rapid retrieval, triage, strategic judgment, emotional regulation, and team communication. Treating these layers as identical encourages introductory courses to assess capabilities they have not yet taught.

The curriculum presented here uses backward design to put enduring algorithmic practices before topic accumulation and mastery before consequential speed. CP1 establishes a broad repertoire and a language for recognizing structure. CP2 increases ambiguity, integration, and transfer. CP3 restores competition as an explicitly taught and assessed capstone environment. Across the sequence, online-judge feedback is supplemented by human evidence of reasoning, collaborative analysis is linked to individual accountability, and metacognition is treated as part of problem-solving competence rather than an optional reflection exercise.

The title's ``removal'' is therefore temporary and strategic: competition is delayed until it can function as an aligned whole-task assessment rather than an accidental prerequisite. The broader proposition is applicable beyond CP: when an expert practice combines domain knowledge with high-pressure performance, educators should identify which components novices must first learn in protected conditions and which should later be recombined authentically. The appended design tools make that proposition inspectable and adaptable. The proposed evaluation agenda makes it falsifiable.

\section*{Generative AI Use Disclosure}
OpenAI ChatGPT (GPT-5.6 Pro, accessed July--August 2026) assisted with an early working draft and provided targeted suggestions on selected passages and \LaTeX{} source during revision. The author conceived the curriculum and article, developed the paper's argument, supplied and interpreted all underlying materials, wrote or substantially revised the majority of the submitted text, made all substantive decisions, and verified the claims, citations, and final wording. The author accepts full responsibility for the manuscript.

\section*{Data and Materials Availability}
This article analyzes curriculum artifacts rather than a student dataset. The appendices provide reusable design instruments and implementation protocols. \Cref{app:cap} reproduces and lightly adapts the CAP curriculum-design framework with permission from its author. The underlying local syllabi, proposals, and working documents are cited as unpublished artifacts; please contact the corresponding author to request originals. No student records, submissions, interview data, or generative AI chat histories were used in this paper.

\section*{Acknowledgments}

The curriculum-design process originated in Purdue University's ENE 50600, Content, Assessment and Pedagogy, taught by Ruth A. Streveler. The author thanks Professor Streveler for permission to reproduce and adapt the CAP project framework, and thanks the supervising faculty, teaching assistants, and students whose questions and implementation experiences informed successive revisions of the course.

\appendices
\crefalias{section}{appendix} %

\section{Content--Assessment--Pedagogy Curriculum Design Framework} \label{app:cap}

This appendix reproduces and lightly adapts, with permission from \CAPauthor{}, the curriculum-design project framework used in \CAPcourse{}, \cite{streveler2021capsyllabus}. Requirements specific to the original graduate assignment, such as explaining the project's value to the designer's career or using a prescribed document format, have been removed. The sequence and mode of reasoning are preserved. That sequence is substantive: the designer first analyzes the domain, then narrows from the domain to course outcomes, integrates a learner-centered analysis, specifies evidence, designs assessment and pedagogy, and finally demonstrates alignment.

\subsection{Introduction: Setting and Design Motivation}

Begin by defining the instructional setting before selecting content.
\begin{enumerate}[label=\textbf{1.\arabic*.}]
    \item Give the project a title that identifies the course, module, workshop, or other instructional unit and its intended learners.
    \item Describe salient characteristics of the institution or sponsoring organization, including the role of the instruction in the larger program, typical enrollment, and relevant disciplinary culture.
    \item Identify external constraints on the design. Examples include required texts or platforms, prerequisites, credit hours, calendar and contact time, downstream courses or events, staffing, facilities, and institutional or accreditation requirements.
    \item Identify contextual conditions that can affect learning or participation, including access to technology and educational resources, disability, employment, commuting, caregiving, financial or food insecurity, linguistic background, marginalization, and prior educational opportunity.
    \item Explain the designer's motivation and relevant expertise. Separate disciplinary expertise, educational evidence, and local implementation experience so readers can see the basis for later decisions.
    \item State the design problem and the intended scope of the redesign. Identify both the desired change and important non-goals.
\end{enumerate}

\subsection{Content: Knowledge-Centered Analysis}

\paragraph{Look first at the domain or subject area and ask}
\begin{enumerate}[label=\textbf{2.1.\arabic*.}]
    \item What are the big ideas of the target domain? Write each as a complete statement rather than as a topic label.
    \item What are the guiding concepts of the target domain? State the relationships, models, conditions, or modes of reasoning that allow learners to act on the big ideas.
    \item What are the essential questions for the target content? These should be recurring questions that experts ask and that learners should eventually ask themselves.
\end{enumerate}

\paragraph{Now narrow from the domain to the outcomes of the particular course and ask}
\begin{enumerate}[label=\textbf{2.2.\arabic*.}]
    \item What are the enduring outcomes: the capabilities learners should retain and use after many details have been forgotten?
    \item What are the important-to-know outcomes: concepts, procedures, canonical cases, and terminology that enable the enduring outcomes?
    \item What are the good-to-be-familiar-with or supplemental outcomes? This category is optional and should not silently become required evidence of mastery.
\end{enumerate}

This transition prevents the domain from being mistaken for the course. A field may contain hundreds of valued topics, while a course must select a coherent subset in service of a smaller number of durable capabilities. In CP, for example, recognizing when shortest-path reasoning applies is an enduring practice; implementing one canonical shortest-path algorithm is important-to-know knowledge that supports it.

\paragraph{Tie the domain and course analyses together}
\begin{enumerate}[label=\textbf{2.3.\arabic*.}]
    \item Create a concept map that illustrates relationships among the guiding concepts and shows how they align with enduring and important-to-know outcomes. Distinguish prerequisite, enabling, contrasting, and transfer relationships rather than drawing only topic adjacency.
    \item Describe the concept map in words. Explain why the relationships justify the proposed scope and sequence.
\end{enumerate}

\subsection{Content: Learner-Centered Analysis}

Analyze the intended learners independently of the content list.
\begin{enumerate}[label=\textbf{2.4.\arabic*.}]
    \item What prior knowledge are learners likely to possess in the target domain? Include variation, fragile prerequisites, and knowledge that may be available but inert.
    \item What age or developmental characteristics are relevant? What emotional, neurological, physical, or linguistic conditions may affect participation or representation of knowledge?
    \item What external factors may affect learning, such as illness, poverty, housing or food insecurity, discrimination or marginalization, work, commuting, or caregiving?
    \item What other characteristics of the learner population require attention, including different reasons for enrolling, prior experience, or accelerated pathways?
\end{enumerate}

\subsection{Intersection of the Knowledge- and Learner-Centered Analyses}

Bring the first two analyses together before writing assessments.
\begin{enumerate}[label=\textbf{2.5.\arabic*.}]
    \item What difficult concepts, practices, or misconceptions have been identified in the target domain for these learners? Difficulty should name a learning mechanism rather than merely an advanced topic.
    \item Why are those concepts difficult? State a theory of difficulty. Possibilities include tacit expert knowledge, ritual knowledge, inert knowledge, foreign or counterintuitive knowledge, excessive element interactivity, representational inflexibility, fragile prerequisites, motivation, or another mechanism supported by evidence \cite{perkins2009whole} (pp. 89-105).
    \item What observable learner behavior would indicate that the difficulty or misconception has been productively addressed?
    \item Which differences among learners call for scaffolding, choice, alternate representations, or accelerated pathways?
\end{enumerate}

\subsection{Evidence for Content Decisions}

Explain the evidence used to select the big ideas, enduring outcomes, difficult concepts, and major sequence decisions. The original framework identifies five categories, ordered from strongest to more local forms of warrant for the purposes of the design exercise:
\begin{enumerate}
    \item discipline-based education research;
    \item general research on learning (e.g. \cite{svinicki2004learning});
    \item expert disciplinary opinion, including textbooks and statements about what matters in the field;
    \item syllabi or documented designs of comparable courses; and
    \item the designer's own experience learning or teaching the content.
\end{enumerate}

A useful evidence table has one row per consequential claim and records the claim, evidence category, sources, applicability to the intended learners, uncertainty or disagreement, and resulting design decision. Local experience can motivate a design choice, but labeling it as such prevents it from being confused with independently established evidence.

\subsection{Assessment}

\begin{enumerate}[label=\textbf{3.\arabic*.}]
    \item Learning objectives
    \begin{enumerate}[label=\textbf{3.1.\arabic*.}]
        \item Create a list of measurable learning objectives for the instruction. Use observable verbs and preserve conceptual, procedural, metacognitive, and, when relevant, interpersonal dimensions.
        \item Mark the objectives that directly measure whether students have learned the difficult concepts or overcome the misconceptions identified above.
        \item Describe how each learning objective aligns with the enduring and important-to-know outcomes.
        \item Create a table showing how the objectives fit a chosen taxonomy of learning, such as the revised Bloom taxonomy \cite{anderson2001taxonomy}, Fink \cite{fink2003self}, DOK \cite{webb2002depth}, etc.
    \end{enumerate}
    \item Create an assessment worksheet for at least three objectives that measure enduring outcomes, difficult concepts, or misconceptions. For each, specify the conditions of performance, the student product or action, criteria for success, opportunities for formative feedback, and threats to valid interpretation.
    \item Select one consequential assessment and explain how it meets six authentic-assessment criteria: realistic context, judgment and innovation, doing the subject, relevant real-world or disciplinary constraints, use of a repertoire of knowledge and skill, and opportunities for actionable feedback \cite{hansen2011idea}.
    \item Find or create a rubric for the authentic assessment and indicate the proficiency level required for passing.
    \item Build an alignment matrix showing which assessments provide primary and secondary evidence for each objective. Reconsider graded work that has no clear outcome function.
\end{enumerate}

\subsection{Pedagogy}

\begin{enumerate}[label=\textbf{4.\arabic*.}]
    \item Include a syllabus that states the enduring and important-to-know outcomes, learning objectives, grading standards, criteria for each graded assignment, teaching methods and their rationale (what the class will be like), instructor commitments, preparation and participation expectations, advice for success (including clear prep and behavior for class, how to read and approach class materials, and how to study), and a meeting-by-meeting schedule.
    \item Include a detailed lesson plan for one class period that addresses a named difficult concept or misconception. State the learning objectives, identify the difficulty, specify constructive or interactive activities, anticipate responses and feedback moves, and provide a timeline.

    \item Explain how the instruction addresses the principles listed in \emph{Making Learning Whole} \cite{perkins2009whole}. For each principle, provide a concrete instructional example and the specified connection to Perkins, Svinicki, or the cooperative-learning materials \cite{perkins2009whole,svinicki2004learning,johnson2020cooperativelearning}:
    \begin{enumerate}[label=\textbf{4.3.\arabic*.},itemsep=2pt,labelindent=-1.5em]
        \item \emph{Play the Whole Game}: provide a junior version of the authentic practice.
        \begin{enumerate}[label=\textbf{4.3.\arabic{enumii}.\arabic*.},labelindent=-2.5em,leftmargin=*,nosep]
            \item Provide at least one concrete example of where this principle will be used in the instruction.
            \item Tie to a strategy suggested by Perkins or Svinicki.
        \end{enumerate}

        \item \emph{Make the Game Worth Playing}: make relevance, choice, competence, and community visible.
        \begin{enumerate}[label=\textbf{4.3.\arabic{enumii}.\arabic*.},labelindent=-2.5em,leftmargin=*,nosep]
            \item Provide at least one concrete example of how the instruction will foster motivation.
            \item Tie to a strategy suggested by Perkins or Svinicki.
            \item Tie to a theory of motivation discussed by Perkins or Svinicki.
        \end{enumerate}

        \item \emph{Work on the Hard Parts}: give difficult subskills focused practice while retaining the whole-task context.
        \begin{enumerate}[label=\textbf{4.3.\arabic{enumii}.\arabic*.},labelindent=-2.5em,leftmargin=*,nosep]
            \item Provide at least one concrete example of how the instruction will ``embrace the hard parts.''
            \item Tie to a strategy suggested by Perkins or Svinicki.
            \item Link to a specific difficult concept or misconception identified in Item~\textbf{2.5.1}.
        \end{enumerate}

        \item \emph{Play Out of Town}: promote transfer across contexts and representations.
        \begin{enumerate}[label=\textbf{4.3.\arabic{enumii}.\arabic*.},labelindent=-2.5em,leftmargin=*,nosep]
            \item Provide at least one concrete example of how the instruction will promote transfer.
            \item Tie to a strategy suggested by Perkins or Svinicki.
        \end{enumerate}

        \item \emph{Uncover the Hidden Game}: make tacit expert cues, judgments, and monitoring explicit.
        \begin{enumerate}[label=\textbf{4.3.\arabic{enumii}.\arabic*.},labelindent=-2.5em,leftmargin=*,nosep]
            \item Be explicit about the ``hidden game'' being uncovered.
            \item Provide at least one concrete example of how the instruction will uncover that hidden game.
            \item Tie to a strategy suggested by Perkins or Svinicki.
        \end{enumerate}

        \item \emph{Learn from the Team}: create interactive work in which learners co-construct and critique reasoning.
        \begin{enumerate}[label=\textbf{4.3.\arabic{enumii}.\arabic*.},labelindent=-2.5em,leftmargin=*,nosep]
            \item Provide at least one concrete example of an interactive activity that will help learners understand an enduring outcome.
            \item Using examples from Karl Smith's cooperative-learning reading or video \cite{johnson2020cooperativelearning}, explain how the activity will promote co-construction of knowledge among groups of learners \cite{johnson1998cooperative}.
        \end{enumerate}

        \item \emph{Learn the Game of Learning}: promote planning, monitoring, reflection, and metacognition.
        \begin{enumerate}[label=\textbf{4.3.\arabic{enumii}.\arabic*.},labelindent=-2.5em,leftmargin=*,nosep]
            \item Provide a concrete example of how the instruction will promote metacognition in learners.
            \item Tie to a strategy suggested by Perkins or Svinicki.
        \end{enumerate}
    \end{enumerate}

    \item For each major objective, identify deliberate, distributed practice and feedback opportunities rather than relying on one-time exposure \cite{ericsson1993practice,cepeda2006distributed}.
    \item Distinguish passive, active, constructive, and interactive activity and justify where interaction should improve knowledge construction \cite{chi2009active,chiwylie2014icap}.
    \item State how scaffolds will fade: what the instructor initially models, what prompts learners receive, and what learners must eventually do independently.
\end{enumerate}

\subsection{Alignment Statement}

Conclude with a narrative explaining how content, assessment, and pedagogy operate as one design. One productive structure is to select the most important enduring outcome and trace it through the course: why it matters, which assessment provides acceptable evidence, where learners engage in deliberate distributed practice, how feedback changes later performance, and how the course avoids assessing capabilities it has not taught.

Then conduct a bidirectional audit:
\begin{itemize}
    \item For every outcome, where is it practiced, assessed, and revisited?
    \item For every assessment, which outcome warrants its inclusion?
    \item For every major activity, what learner thinking is expected and what evidence indicates that it occurred?
    \item For every prerequisite, how is readiness established or repaired?
    \item For every policy, what educational function justifies its burden or restriction?
\end{itemize}

\subsection{Extension for an Implemented Curriculum}

The original CAP project culminates in the alignment statement. An implemented curriculum can add a design-research extension: state the design conjectures and plausible alternatives, select outcome and feasibility measures proportional to the intended claims, record major adaptations and contextual changes, separate course assessment from research participation, and document what changes in the next version. \Cref{sec:evaluation,app:evaluation} provide a tiered example of that extension.

\section{Illustrative CP1 Schedule and Outcome Mapping} \label{app:schedule}

\Cref{tab:schedule} translates the Spring 2026 topic sequence into an outcome-centered schedule. It is illustrative rather than a specific recommendation: instructors should change problems and pacing while preserving prerequisite logic and cumulative practice.

\begin{table*}[t]
    \centering
    \caption{Illustrative twelve-topic CP1 schedule. Outcome labels refer to O1--O6 in \Cref{sec:content}.}
    \label{tab:schedule}
    \scriptsize
    \begin{tabularx}{\textwidth}{P{0.065\textwidth} P{0.18\textwidth} Y Y P{0.15\textwidth}}
        \toprule
        \textbf{Topic} & \textbf{Content emphasis} & \textbf{Reasoning emphasis} & \textbf{Representative in-class evidence} & \textbf{Primary outcomes} \\
        \midrule
        0 & Course introduction; constraints; overflow; input/output; problem-solving cycle & Translate statement and samples; connect bounds to candidate complexity; establish brute-force-first routine & Annotate a small problem and reject two infeasible designs & O1, O3, O6 \\
        1 & Useful data structures; sorting; complete search & Choose a representation based on required operations; enumerate without omission or duplication & Compare array, map, heap, stack, and bitmask representations & O1-O4 \\
        2 & Greedy algorithms & Identify local choice; formulate exchange or stay-ahead justification; construct counterexamples & Debate plausible rules and repair a failed greedy argument & O1-O3, O6 \\
        3 & Divide and conquer & Identify independent subproblems and combine step; derive recurrence informally & Contrast recursive partitioning with exhaustive search & O1-O4 \\
        4 & Bisection/binary search the answer; meet in the middle & Detect monotone feasibility; separate answer space from construction; split exponential search & Prove monotonicity and design the feasibility predicate (``given the answer, can you tell me if it's correct efficiently'') & O1-O4, O6 \\
        5 & Dynamic programming foundations & Define state, transition, base cases, evaluation order, and invariant meaning & Convert recursion to memoization and tabulation; explain state sufficiency & O1-O4 \\
        6 & Dynamic-programming applications; prefix sums; two pointers & Compare overlapping-subproblem and incremental-window models; identify reusable partial computation & Diagnose an over-specified DP state and derive a linear alternative & O1-O5 \\
        7 & Advanced introductory DP and mixed paradigms & Trace reconstruction; test state boundaries; integrate a secondary technique & Explain and modify a noncanonical DP solution & O2-O6 \\
        8 & Graph review; DFS/BFS; connected and strongly connected components & Choose graph representation; formulate reachability invariants; distinguish directed connectivity notions & Draw graph from story, predict traversal, and justify component algorithm & O1-O5 \\
        9 & Flood fill; topological reasoning & Convert spatial problems to graphs; recognize partial order and cycle implications & Compare BFS/DFS flood fill; simulate Kahn's algorithm & O1-O5 \\

        10 & Shortest paths; CP2 preview (MSTs, disjoint sets/union find) & Select BFS, Dijkstra's algorithm, or another shortest-path method from edge assumptions; distinguish shortest-path and spanning-tree objectives; connect greedy edge selection to disjoint-set cycle detection & Reject Dijkstra under violated assumptions; trace Kruskal's algorithm with union-find and justify each accepted or rejected edge & O1-O6 \\

        11 & Random/mixed problem solving & Classify without a topic label; integrate prior schemas; monitor strategy and uncertainty & Individual analysis, peer critique, implementation plan, retrospective & O1-O6 \\
        \bottomrule
    \end{tabularx}
\end{table*}

A weekly problem set should include variation in surface context and difficulty. One task can be a highly scaffolded canonical case, one can require transfer within the week's family, one can contrast a common misconception, and one can integrate a prior topic. The breadth gate could be displayed as an outcome dashboard so that students can see which families remain unevidenced before the final weeks.

\section{Sampled Oral Interview Protocol and Analytic Rubric} \label{app:interview}

Continuing from the discussion in \Cref{sec:assessment:interviews}, this protocol supplements accepted code with a brief sample of reconstructable understanding. Students should receive the prompt categories and rubric in advance. The assessor selects one previously accepted problem, with brief scratch work allowed but the submitted code hidden. A normal interview lasts 5-15 minutes.

\subsection{Core Interview Sequence}
\begin{enumerate}[label=\textbf{I\arabic*.}]
    \item \textbf{Comprehend.} ``What does the problem ask you to compute? Explain one sample input and output in your own words.''
    \item \textbf{Baseline and design.} ``What is a straightforward or brute-force approach? Explain the approach you submitted, the decisive insight, and the relevant CP paradigm or terminology.''
    \item \textbf{Analyze.} ``What are the time and space bounds, and why are they sufficient for the stated constraints?''
    \item \textbf{Targeted probe, only when needed.} Ask one brief question that resolves an ambiguity in the preceding answers, such as ``What happens next on this small input?'', ``Which assumption makes that step valid?'', or ``Where did your original attempt fail?''
\end{enumerate}

A custom test, full correctness proof, or transfer modification can be valuable in a longer oral examination, but requiring all of them would exceed the purpose and time budget of this sampled interview. The assessor should use neutral probes before offering content hints and give the student the space to answer. Minor syntax and naming differences are irrelevant; the target is the student's model of the problem and solution. The full rubric can be seen in \Cref{tab:oralrubric}.

\begin{table*}[t]
    \centering
    \caption{Compact analytic rubric for a sampled oral code interview.}
    \label{tab:oralrubric}
    \scriptsize
    \begin{tabularx}{\textwidth}{P{0.17\textwidth} Y Y Y}
        \toprule
        \textbf{Dimension} & \textbf{0 -- Insufficient evidence} & \textbf{1 -- Partial evidence} & \textbf{2 -- Sufficient evidence} \\
        \midrule
        Problem comprehension and sample & Cannot accurately restate the task or explain the sample; omits a condition that changes the problem & States the main goal but needs substantial prompting or misses a consequential detail & Accurately explains the goal, relevant constraints, and sample in the student's own words \\

        Algorithmic reasoning and CP vocabulary & Cannot verbally reconstruct the submitted approach or gives reasoning incompatible with the accepted behavior & Identifies the main technique but leaves important gaps, assumptions, or the decisive insight unexplained & Reconstructs the baseline and accepted design, identifies the key insight, and uses relevant course terminology coherently \\

        Complexity and constraint linkage & Cannot state meaningful bounds or gives bounds inconsistent with the approach & Gives approximately correct bounds but cannot derive or connect them convincingly to the input limits & Derives time and space bounds and explains why the implementation is feasible under the stated constraints \\
        \bottomrule
    \end{tabularx}
\end{table*}

A practical passing profile is at least 5 of 6 points with no zero. When the interview does not establish that profile, the sampled solution loses credit and the submission is reviewed for possible plagiarism or unauthorized assistance under the ordinary institutional process. The review is exploratory and may clear the submission; the interview score itself is not a misconduct finding. A short follow-up interview is appropriate when the initial result is genuinely ambiguous because of communication, prompt interpretation, or an accommodation issue, especially when the interview is performed by instructional staff other than the course instructor. Interviewers can calibrate efficiently by discussing several scripted examples before the first week of interviews, or by shadowing and being observed during the first round.

\section{Teaching-Assistant Coaching Protocol and Honors Pathways} \label{app:ta}
\subsection{A Diagnose-Elicit-Focus-Scaffold-Fade-Verify Protocol}
Office hours should be designed as coached problem solving rather than either answer withholding or answer delivery. The following protocol operationalizes cognitive-apprenticeship and scaffolding principles \cite{collins1989apprenticeship,wood1976scaffolding}.

\begin{enumerate}[label=\textbf{T\arabic*.}]
    \item \textbf{Diagnose the current model.} Ask the student to restate the task, explain a sample, state constraints, and describe what they have tried. Inspect whether the bottleneck is comprehension, strategy, proof, complexity, implementation, or debugging.
    \item \textbf{Elicit existing resources.} Ask for the brute-force method, relevant paradigms, similar prior problems, and data structures whose operations match the task. Require the student to externalize a diagram, state table, or small trace when useful.
    \item \textbf{Focus attention.} Direct the student to one discriminating feature or contradiction without naming the complete solution: ``What changes monotonically?'', ``Which work is repeated?'', ``What would make this choice unsafe?'', ``Which edge assumption does Dijkstra require?''
    \item \textbf{Scaffold one step.} If attention prompts fail, provide a partial representation, subproblem, invariant, counterexample, or next debugging experiment. Avoid supplying code unless the learning objective is code reading or repair.
    \item \textbf{Fade.} Return control immediately: ask the student to complete the reasoning, predict the next step, or implement the hinted component.
    \item \textbf{Verify understanding.} Before the student leaves, ask for a summary, complexity argument, and a new test or variation. A student who can repeat the hint but cannot transfer it has not yet demonstrated understanding.
    \item \textbf{Document recurring barriers.} Record anonymized patterns for course improvement. Multiple students needing the same hint may indicate a problem-selection, lecture, or prerequisite issue rather than individual deficiency.
\end{enumerate}

TA calibration meetings can use short cases: an incorrect greedy rule, an overlarge DP state, a Dijkstra solution with negative edges, an off-by-one prefix sum, and a student who has accepted code but cannot explain it. Staff should compare the minimum next hint that keeps the student cognitively active. This practice reduces the tendency for some assistants to over-help while others valorize unproductive struggle.

\subsection{Honors Pathway 1: Authoring and Validating a Problem}
Problem creation is the preferred honors pathway because it requires a learner to reverse the usual perspective. The student must design not only a solvable task but also a specification, intended solution, proof, data limits, tests, and difficulty experience. A complete project should include:

\begin{itemize}
    \item a clear problem statement with input/output specification, constraints, and samples;
    \item an intended algorithm and at least one Time Limit Exceeded solution along with a plausible wrong or suboptimal approach;
    \item a correctness argument and time/space analysis;
    \item a validator, reference solution, systematic large-test-case generator, and hand-curated adversarial test suite;
    \item a rationale for constraints and expected difficulty;
    \item evidence from at least two peer solvers, including observed misconceptions and revisions;
    \item a postmortem mapping the task to CP1/CP2 outcomes and explaining what the author learned about problem setting.
\end{itemize}

This pathway authentically integrates specification, algorithm design, debugging, assessment, and perspective taking. It should be evaluated with milestones so that an invalid or miscalibrated problem can be repaired rather than discovered at the end. \textbf{A template ``problem skeleton'' and assignment milestones are available from the corresponding author upon request.}

\subsection{Honors Pathway 2: Advanced Transfer Portfolio}
An alternative is a portfolio of advanced or multi-technique problems. Additional quantity alone is not sufficient. The portfolio should require diversity, explanation, and transfer: for each selected problem, the student submits the accepted solution, a design memo, correctness and complexity arguments, a custom test suite, and a comparison with at least one alternative approach. A final synthesis identifies recurring cues, failed strategies, and connections among paradigms. This pathway is easier to administer but less generative than problem authoring; it should therefore require deeper evidence than ``solve more problems.''

\section{Extended Evaluation Framework} \label{app:evaluation}

Continuing the discussion from \Cref{sec:evaluation:extended}, this appendix collects measures that support a full program of research but are not required for every offering. They can be added modularly to the minimum-viable designs in \Cref{tab:minstudy}.

The full evidence matrix can be seen in \Cref{tab:evaluation}.

\begin{table*}[t]
    \centering
    \caption{Extended evidence plan for evaluating the curriculum sequence.}
    \label{tab:evaluation}
    \scriptsize
    \begin{tabularx}{\textwidth}{P{0.13\textwidth} P{0.18\textwidth} Y Y P{0.15\textwidth}}
        \toprule
        \textbf{Construct} & \textbf{Primary measure} & \textbf{Evidence collected} & \textbf{Interpretive control} & \textbf{Analysis unit} \\
        \midrule
        Classification and decomposition & Novel written or oral transfer task scored with an analytic rubric & Problem-family hypotheses, structural cues, brute-force baseline, subproblem map, counterexamples & Use unfamiliar surface contexts and common scoring criteria & Student $\times$ task $\times$ time \\
        Complexity and feasibility & Constraint-linked algorithm analysis & Time/space derivation, assumptions, threshold comparison, rejection of infeasible designs & Include plausible distractor algorithms and score the reasoning path & Student $\times$ task \\
        Implementation and debugging & Common programming task plus a targeted test-generation prompt & Accepted code, attempt trace, verdict sequence, custom tests, defect explanation & Interpret submission counts with code changes or explanation & Student $\times$ problem \\
        Explanation and authorship & Sampled oral interview & Problem restatement, baseline, key insight, paradigm, and complexity linkage & Use the compact rubric and a targeted follow-up only when needed & Student $\times$ interview \\
        Breadth and transfer & Topic-tagged portfolio plus delayed near- and far-transfer tasks & Topic families demonstrated, multi-technique solutions, delayed retention & Treat the gate as coverage evidence rather than complete mastery & Student $\times$ outcome family \\
        Contest performance & Repeated individual and team contests in CP3 & Time to first acceptance, solves, penalties, problem order, switching decisions, handoffs, and idle time & Compare repeated events and report absolute indicators alongside rank & Student/team $\times$ contest \\
        Motivation and participation & Brief repeated surveys plus selected interviews or observations & Self-efficacy, goals, belonging, perceived climate, intention to continue, discussion patterns & Collect only constructs needed for an explicit research question & Student $\times$ time \\
        Feasibility and fidelity & Instructor and TA logs & Preparation time, interview minutes, office-hour demand, activity completion, major adaptations & Use a short recurring log rather than continuous observation & Offering $\times$ week \\
        \bottomrule
    \end{tabularx}
\end{table*}

\subsection{Motivation and Participation Module}

RQ3 requires evidence beyond aggregate performance. A manageable version uses a brief beginning/end survey of algorithmic self-efficacy, belonging, and intention to continue, together with prior CP experience. A fuller study can add mastery- and performance-goal orientations, perceived autonomy support, interviews, or structured observations of who initiates ideas and who implements. Experienced students should be included explicitly because they may prefer early competition or perceive introductory scaffolds as slow. Optional contests, challenge problems, instructor-approved advanced placement, and repeatable CP3 participation provide plausible mechanisms for preserving their engagement \cite{ryandeci2020sdt,reeve2009autonomy}.

Subgroup analyses should match the sample size and the theory being tested. Prior CP experience is the highest-priority grouping variable because the curriculum is designed for mixed-experience cohorts. Other demographic or life-context variables should be collected only when the study has a clear analytic use and sufficient protection against small-cell disclosure.

\subsection{Fidelity, Cost, and Sustainability Module}

The most useful feasibility record is a brief weekly log of resources that vary with the design: problem-curation time, office-hour demand, interview minutes, plagiarism-review time, TA calibration, and any adaptation from the planned activity. For the oral interviews, ten students averaging roughly 10 minutes each imply about 1.7 assessor-hours per week before unusual follow-up. Recording this cost makes it possible to distinguish a sound educational mechanism from an implementation that cannot be sustained at another staffing level.

Fidelity need not become an exhaustive observation protocol. A small number of high-leverage checks are sufficient: whether individual thinking preceded peer discussion, whether students had to justify rather than merely name an algorithm, whether staff diagnosed before hinting, whether every student was sampled for an interview, and whether the intended exception processes for attendance-linked materials were applied. Major adaptations should be logged because they help explain outcome differences across offerings \cite{dbr2003design,mckenney2018edr}.

\subsection{Research Ethics and Data Governance Module}

A future study of enrolled students should obtain the institutional determination appropriate to the planned data and should separate voluntary research participation from course grading wherever possible. Audio, screen recordings, code histories, and GenAI chat logs require source-specific decisions about consent, access, retention, and de-identification. The speaking-time diagnostic described in \Cref{sec:pedagogy} is a course-management practice in the present design and is not analyzed as research data in this article.

Assessment-security review should remain distinct from research analysis. A weak oral interview can prompt examination of a submission under the course policy, while any research dataset should be governed by its approved purpose. A concise data plan should identify who can access each source, how identifiers are removed or separated, how long records are retained, and how small groups are protected in reporting.

\bibliographystyle{IEEEtran}
\bibliography{references}

\end{document}